\expandafter\ifx\csname pdfoutput\endcsname\relax
	\else
	\pdfoutput=0
	\fi

\documentclass[letter,12pt]{article}
\usepackage{jheppub}
\usepackage{amsmath,amssymb}
\usepackage{caption,subcaption,float}

\usepackage{pstricks,multido}
\psset{unit=1cm,linewidth=1.2pt}
\newrgbcolor{paleblue}{0.6 0.6 1}
\newrgbcolor{palepink}{1 0.6 0.6}
\newrgbcolor{paleyellow}{1 1 0.6}
\newrgbcolor{darkgreen}{0 0.5 0}
\newrgbcolor{orange}{1 0.5 0}
\newrgbcolor{purple}{0.8 0 1}

\usepackage{chngcntr}
\counterwithin{figure}{section}
\counterwithin{equation}{section}

\DeclareSymbolFont{AMSa}{U}{msa}{m}{n}
\DeclareSymbolFont{AMSb}{U}{msb}{m}{n}
\let\Box\relax
\DeclareMathSymbol{\Box}{\mathord}{AMSa}{"03}

\def\IZ{{\mathbb Z}}
\def\IR{{\mathbb R}}
\def\IC{{\mathbb C}}

\def\vev#1{\left\langle{#1}\right\rangle}
\def\Re{\mathop{\rm Re}}
\def\Im{\mathop{\rm Im}}
\def\tr{\mathop{\rm tr}\nolimits}

\newcommand{\be}{\begin{equation}}
\newcommand{\ee}{\end{equation}}

\title{Deconstructing the $E_0$ SCFT\\ to Solve the Orbifold Paradox\\
	of the Heterotic M Theory}

\author{Jacob Claussen and Vadim Kaplunovsky}
\affiliation{Theory Group, The University of Texas at Austin\\
	Austin, TX, 78712, USA}
\emailAdd{jcclauss@physics.utexas.edu}
\emailAdd{vadim@physics.utexas.edu}

\abstract{%
Many heterotic orbifold models have massless twisted-sector particles with
simultaneous $E8_1$ and $E8_2$ charges.
In the strong-coupling M--theory dual of the heterotic string, this poses a paradox:
Since the $E8_1$ and $E8_2$ live at opposite ends of the $x^{10}$ dimension,
where could a massless particle with both types of charges possibly live?
To key to this question are the 5D SCFTs living at the orbifold fixed planes
going through the bulk of the M theory.
We use dimensional deconstruction to understand how such a 5D SCFT (specifically,
the $E_0$ SCFT at the $\IZ_3$ fixed point) works at the superconformal point
(rather that at the Coulomb branch) and how it interacts with the boundaries of the
$x^{10}$.
We find that the massless twisted states are not localized in the $x^{10}$.
Instead, they are non-local meson-like composite particles comprised of a quark
living at one boundary of the $x^{10}$, an antiquark living at the other boundary,
and the string of strongly-interacting 5D gluons connecting the quark to the antiquark.
}

\preprint{UTTG--11--16}

\begin{document}
\maketitle


%

\section{Introduction and Summary}
Heterotic orbifolds 
\cite{Dixon:1985jw,Dixon:1986jc}
are among the oldest and best-known types of string models.
Unlike the smooth Calabi--Yau compactifications of the $E8_1\times E8_2$ heterotic string,
many $T^6/({\rm discrete\ symmetry\ \mit D})$ compactifications have massless particles
charged under (unbroken subgroups of) both $E8_1$ and $E8_2$.
Such particles are perfectly normal from the
perturbative heterotic string theory point of view,
but in the strong-coupling dual of the heterotic string --- the Ho\v{r}ava--Witten
theory \cite{HoravaWitten95,HoravaWitten96} --- they pose a paradox.
Indeed, the Ho\v{r}ava--Witten theory is M theory whose eleventh dimension is
a finite interval with boundaries; the supergravity lives in the 11D bulk,
while the $E8_1$ and $E8_2$ SYM theories live on the 10D boundaries
at the opposite ends of the $x^{10}$.
In this setup, massless particles with both $E8_1$ and $E8_2$ charges raise a paradox:
where in the $x^{10}$ can they possibly live?

In the heterotic string theory, the particles in question belong to the {\it twisted sectors}
of the world-sheet Hilbert space, where the string does not close on the ${\bf R}^{1,3}\times T^6$
but closes modulo the discrete symmetry $D$:
\be
X^I(\sigma=2\pi)\ =\ {\cal D}^I_{\,\,J}X^J(\sigma=0),\quad
{\cal D}\in D,\quad {\cal D}\neq1.
\ee
The massless states in a twisted sector obtain from a string which begins and ends on
a {\it fixed point} of the discrete symmetry $D$.
Classically, this allows for an arbitrarily short length of the string loop and hence
arbitrarily low energy; this does not guarantee massless particle states in the quantum theory,
but many orbifold models do have massless twisted states.
Also, if action of the symmetry $D$ breaks both $E8_1$ and $E8_2$ gauge symmetry groups,
then all the twisted states --- massless and massive --- are charged under surviving
subgroups of both $E8_1$ and $E8_2$.

The strong-coupling dual of a heterotic orbifold is M theory on
$$
{\bf R}^{1,3}_{\rm Minkowski}\times\bigl(T^6/D\bigr)_{\rm orbifold}\times\rm Interval.
$$
The massless twisted states are localized in the $T^6/D$ dimensions at the fixed points of $D$,
but their locations in the $x^{10}$ dimension is problematic, especially in light of their
$E8_1\times E8_2$ charges.
In the present paper we solve this problem, and the solution turns out to be very different
from the 6D ${\bf R}^{1,5}\times T^4/D$ orbifolds explored in \cite{Vadim99,Vadim01}:
\begin{itemize}
\item[$\star$]
In the 6D orbifolds, the massless twisted states are localized at one end of the $x^{10}$
dimension.
\item[$\star$]
But in the 4D orbifolds, the massless twisted states span the whole $x^{10}$,
from one end to the other end.
\end{itemize}

In both cases, the key to the twisted stated is the non-trivial physics on the
fixed 7D or 5D planes of the discrete symmetry $D$ in the 11D bulk of the M theory.
For the 6D orbifolds, a fixed plane locally looks like ${\bf R}^{1,6}\times({\bf C}^2/D)$,
which in M theory gives rise to the 7D SYM theory on the ${\bf R}^{1,6}$,
or rather on the $R^{1,5}\times\rm Interval$.
At one end of the Interval, this 7D gauge theory locks onto an unbroken subgroup of
the $E8_1$ in a kind of inter-dimensional Higgs mechanism,
\be
\left( G_{\rm 10D}\subset E8_1\right)\times G_{7D}\
\to\ G_{\rm common}\,.
\ee
In the perturbative heterotic string, this $G_{\rm common}$ gauge symmetry
appears to be a subgroup of the $E8_1$, but in the M theory it lives on both
the 10D boundary at $x^{10}=0$ and the 7D fixed planes, and along those fixed planes it
reaches all the way to the other end of the $x^{10}$ where it meets the $E8_2$
gauge group (or rather its surviving subgroups).
Thanks to this meeting of gauge symmetries, the 6D hypermultiplets localized at the intersections
of the fixed planes with the second end of the $x^{10}$ may carry both
the $G_{\rm common}$ and  the $E8_2$ charges --- which in the heterotic limit looks like
the simultaneous $G_{10\rm D}\subset E8_1$ and $E8_2$ charges.

For an example, consider the $T^4/\IZ_2$ orbifold in which the $E8_1$ is broken down to
$E7\times SU2$ while the $E8_2$ is broken down to $SO16$.
This orbifold has 16 fixed points on the $T^4$, each fixed point giving rise
to  massless half-hypermultiplets in the $({\bf 1},{\bf 2};{\bf 16})$ multiplet
of the $E7\times SU2\times SO16$ gauge group.
The simultaneous $SU2\subset E8_1$ and $SO16\subset E8_2$ quantum numbers are
best explained pictorially, see figure~\ref{fig:orbifold6D}:
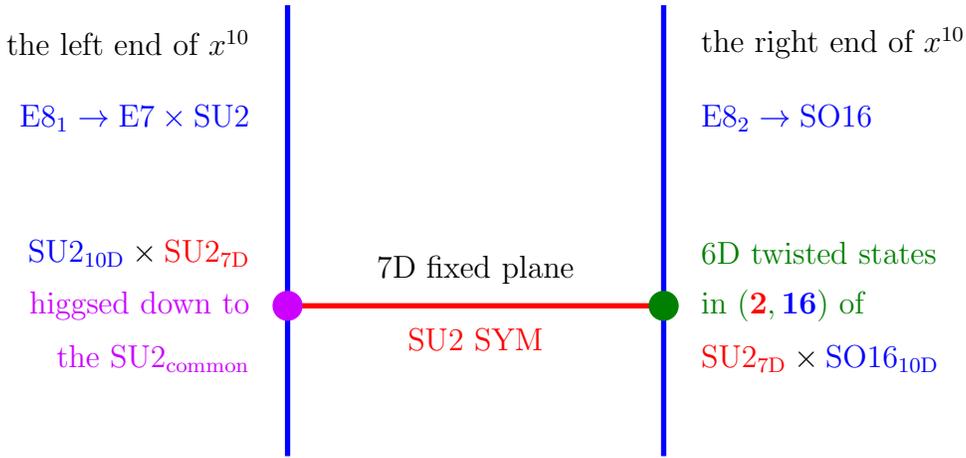
\begin{figure}[H]
\psset{labelsep=0.5}
\begin{pspicture}[shift=-3](-7.5,-2)(+7.5,+4.5)
\psline[linecolor=blue,linewidth=2pt](-2.5,-2)(-2.5,+4)
\psline[linecolor=blue,linewidth=2pt](+2.5,-2)(+2.5,+4)
\psline[linecolor=red,linewidth=2pt](-2.5,0)(+2.5,0)
\uput[u](0,-0.2){7D fixed plane}
\uput[d](0,+0.2){\red SU2 SYM}
\uput[l](-2.5,+3.5){the left end of $x^{10}$}
\uput[r](+2.5,+3.5){the right end of $x^{10}$}
\uput[l](-2.5,+2.5){$\rm\blue E8_1\to E7\times SU2$}
\uput[r](+2.5,+2.5){$\rm\blue E8_2\to SO16$}
\pscircle*[linecolor=darkgreen](+2.5,0){0.2}
\rput[l](+3,+0.7){\darkgreen 6D twisted states}
\rput[l](+3,0){\darkgreen in $({\bf\red 2},{\bf\blue 16})$  of}
\rput[l](+3,-0.7){${\rm\red SU2_{7D}}\times{\blue\rm SO16_{10D}}$}
\pscircle*[linecolor=purple](-2.5,0){0.2}
\rput[r](-3,+0.7){${\rm\blue SU2_{10D}}\times{\rm\red SU2_{7D}}$}
\rput[r](-3,0){\purple higgsed down to}
\rput[r](-3,-0.7){\purple the $\rm SU2_{common}$}
\end{pspicture}
\caption{Twisted states of the 6D $T^4/\IZ_2$ orbifold.}
\label{fig:orbifold6D}
\end{figure}
The massless 6D twisted states live at the intersections of the 7D fixed planes
of the orbifold with the 10D right boundary of the 11D bulk of M theory.
Their gauge quantum numbers are $\bf 16$ of the $SO16\subset E8_2$ living on the
10D right boundary of the $x^{10}$ and $\bf 2$ of the $SU2$ living on the 7D fixed plane;
both gauge symmetries are present at the intersection, so the simultaneous $({\bf 2},{\bf 16})$
quantum numbers are OK.
Note that {\it locally} --- where the massless twisted states live --- the $SU2$ is the
7D gauge theory on the fixed plane rather than the $SU2_{\rm 10D}\subset E8_1$.
However, at the other (left) end of the $x^{10}$, the 7D $SU2$ locks onto
the unbroken $SU2$ subgroup of the 10D $E8_1$, so {\it globally} we have a common $SU2$
which lives both on the 10D left boundary of the $x^{10}$ and on all the 7D fixed planes;
the massless twisted states end up being doublets of this common $SU2$.
From the heterotic string point of view, this common $SU2$ {\it appears to be}
a subgroup of the $E8_1$, but in M theory it is not, and that's what resolves the
paradox of the simultaneous $({\bf 2},{\bf 16})$ charges of the massless twisted states!

For the 4D orbifolds, the situation is more complicated because the fixed planes
--- which locally look like $\IR^{1,4}\times\bigl(\IC^3/D\bigr)$ in the bulk of M theory
--- carry 5D superconformal theories rather than super--Yang--Mills.
For example, the $\IC^3/\IZ_3$ fixed planes of the $T^6/\IZ_3$ orbifold give rise
to the $E_0$ superconformal theories.%
\footnote{%
	The $E_0$ is one of the Morrison--Seiberg $E_n$ superconformal 5D theories
	\cite{MorrisonSeiberg96}.
	In M theory, the $E_n$ SCFTs obtain from Calabi--Yau singularities
	where a del-Pezzo surface collapses to a point.
	The $E_n$'s other than the $E_0$ also obtain in the infinite gauge coupling limit
	of the 5D $SU2$ gauge theories with $n-1$ massless flavors.
	However, the $E_0$ theory is isolated and does not appear in the $g\to\infty$
	limit of any gauge theory.
	}
The $E_0$ theory is poorly understood: All we know is its moduli space
--- which is limited to a 1D Coulomb branch where the $E_0$ reduces to a $U(1)$ SYM with
Chern--Simons level $k=9$ --- and the connection to the moduli spaces of
other 5D theories.
In this paper, we use the flop transition between the Coulomb branches of the $E_0$
and the $SU2$ SYM to {\it deconstruct} \cite{Arkani01}
the fifth dimension of the $E_0$ theory.
In other words, we latticize the $x^{10}$ dimension of the 5D fixed plane, and realize
the $E_0$ theory as a long quiver
\be
\def\nnode{%
	\pscircle(0,0){0.5}
	\rput(0,0){2}
	\psline(0,+0.5)(0,+1)
	\psline(0,-0.5)(0,-1)
	\psline(-0.5,0)(-1.01,0)
	\psline(+0.5,0)(+1.01,0)
	}
\psset{linewidth=1pt}
\begin{pspicture}(-7.5,-1)(+7.5,+1)
\multido{\n=-6+2}{7}{\rput(\n,0){\nnode}}
\psline[linestyle=dotted,linewidth=1.5pt](-7.5,0)(-7,0)
\psline[linestyle=dotted,linewidth=1.5pt](+7.5,0)(+7,0)
\end{pspicture}
\label{LongQuiver}
\ee
of strongly-coupled four-dimensional $SU(2)$ gauge theories.
In the Ho\v{r}ava--Witten orbifold context, the quiver (\ref{LongQuiver}) should have
large but finite length (corresponding to the finite length of the $x^{10}$ dimension),
and there probably should be some extra fields at the two ends of the quiver
corresponding to the 4D fields at the intersections of the 5D fixed plane with the
10D boundaries at the ends of the $x^{10}$.

In this paper, we focus on a particular $T^6/\IZ_3$ orbifold model where the $E8_1$
is broken down to the $E6_1\times SU3_1$ while the $E8_2$ is also broken down to
the $E6_2\times SU3_2$.
The $T^6/\IZ_3$ has 27 fixed points (but no fixed tori), and in this model each fixed point
gives rise to 9 massless chiral multiplet in the
$({\bf 1},{\bf 3};{\bf1},{\bf\bar3})$ multiplet of the unbroken gauge symmetry, that is,
the bi-fundamental multiplet of the two $SU3$ subgroups, one from each $E8$.
The chiral anomaly of these twisted-sector states cancels against the chiral
states in the un-twisted sector, and this cancellation assures that the massless
twisted states stay exactly massless despite any quantum corrections when the
heterotic string becomes strongly coupled.
Therefore, the bi-fundamental $({\bf 3},{\bf\bar3})$ massless particles must exist
in the Ho\v{r}ava--WItten regime of the orbifold, so where in the $x^{10}$ do
these particles live?

To our surprise, we found that these particles are not localized at any particular
place in $x^{10}$ but spread out over the entire $x^{10}$ dimension,
from one boundary to another.
Specifically, the twisted states are meson-like composite particles
comprising a quark at one end of the $x^{10}$,
an antiquark at the other end, and a whole string of 5D gluons connecting
the quark to the antiquark across the whole length of the $x^{10}$,
as shown on figure~\ref{fig:orbifold4d} below:
\begin{figure}[H]
\begin{pspicture}(-7.5,-1.75)(+7.5,+3.5)
\psline[linecolor=blue,linewidth=2pt](-2.5,-1.5)(-2.5,+3.5)
\psline[linecolor=blue,linewidth=2pt](+2.5,-1.5)(+2.5,+3.5)
\psline[linecolor=yellow,linewidth=2pt](-2.5,0)(+2.5,0)
\psline[linecolor=red,linewidth=2pt,linestyle=dashed](-2.5,0)(+2.5,0)
\uput[u](0,+0.1){$\red E_0$ on a 5D fixed plane}
\uput[d](0,-0.1){\red 5D gluons}
\pscircle*[linecolor=darkgreen](-2.5,0){0.2}
\pscircle*[linecolor=darkgreen](+2.5,0){0.2}
\rput[r](-3,+3){left end of the $x^{10}$}
\rput[l](+3,+3){right end of the $x^{10}$}
\rput[r](-3,+2){$\blue\rm E8_1\to E6_1\times SU3_1$}
\rput[l](+3,+2){$\blue\rm E8_2\to E6_2\times SU3_2$}
\rput[r](-3,+0.7){\darkgreen 4D quark}
\rput[r](-3,0){\darkgreen in $({\bf\red q},{\bf\blue 3})$ of}
\rput[r](-3,-0.7){$\rm{\red color}\times{\blue SU3_1}$}
\rput[l](+3,+0.7){\darkgreen 4D antiquark}
\rput[l](+3,0){\darkgreen in $({\bf\red\bar q},{\bf\blue\bar 3})$ of}
\rput[l](+3,-0.7){$\rm{\red color}\times{\blue SU3_2}$}
\end{pspicture}
\caption{Twisted states of a 4D $T^6/\IZ_3$ orbifold}
\label{fig:orbifold4d}
\end{figure}
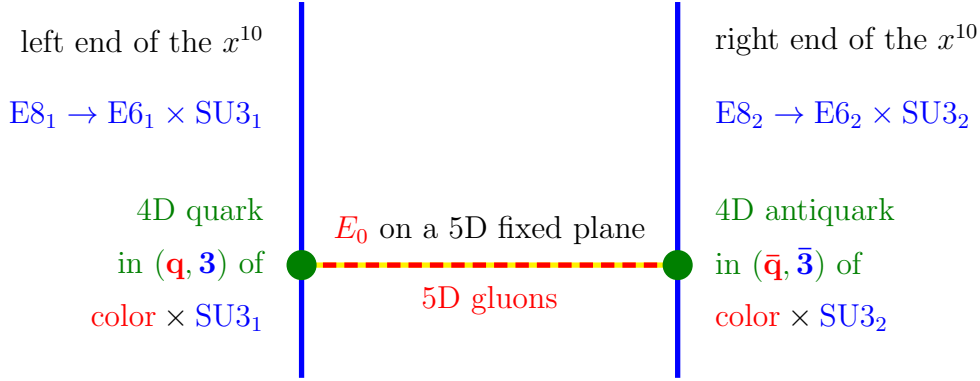
To be precise, in the deconstructed $E_0$ theory on the fixed plane, the twisted
$({\bf 3},{\bf\bar3})$ are composites
\be
{\cal M}_{ij}\ =\  Q_i\Omega_1\Omega_2\cdots\Omega_{N-1}\tilde Q_j
\label{CompositeMesons}
\ee
of the quark $Q_i$ at one end of the quiver, the antiquark $\tilde Q_j$
at the other end, as well as every bifundamental field $\Omega_\ell$ of the quiver.
In the continuum limit, the bifundamental fields becomes components of the 5D gluons
and their superpartners. or at least they become 5D gluons in a 5D gauge theory.
We are not quite sure what exactly do they become in a superconformal theory like $E_0$,
but we call them `gluons' simply because we do not have a better name.

The rest of this paper is the explanation and the justification of the above picture
of the massless twisted states of the $T^6/\IZ_3$ orbifold.
Sections 2 and 3 are introductory:
In section~2 we describe our orbifold model from the heterotic string point of view.
We also consider what happens when a fixed point is `blown up' from both geometrical
and field-theoretical points of view.
In section~3 we discuss the $E_0$ SCFT at the 5D fixed plains in the M-theory dual
of the orbifold.
In \S3.1 we focus on the moduli space of $E_0$ theory in the infinite $4+1$ dimensions
and the flop transitions between the moduli spaces of the $E_0$ and the $SU2$ SYM,
while in \S3.2 we focus on the boundaries of the $x^{10}$
(which acts as the fifth dimension of the $E_0$)
Basically, we summarize  the results of Ganor \& Sonnenschein results \cite{GanorSon02}
about the Coulomb branch of the $E_0$ living on the blown-up fixed plane.

In section 4 we dimensionally deconstruct the $E_0$ theory and explore the
deconstructed theory from the 4D point of view; the \S4 is the core of this paper.
In~\S4.1 we deconstruct the $E_0$ in infinite $4+1$ dimensions:
We start by deconstructing the 5D $SU2$ SYM, then move along the Coulomb branch
of the moduli space across the flop transition to the Coulomb branch of the $E_0$ theory,
and eventually reach the superconformal point of the $E_0$.
We note that in the long quiver limit, there is an abrupt transition
between the semiclassical Higgs regime and the strongly-coupled confinement
regime of the infrared dynamics;
these regimes correspond respectively to the Coulomb branch and the SCFT point
of the 5D $E_0$ theory.
In~\S4.2 we deconstruct the `quarks' and the `antiquarks' at the
boundaries of the $E_0$ theory in the $\IZ_3$ orbifold context,
and we check that the Higgs regime of the resulting quiver theory agrees
with the Ganor--Sonnenschein description of the $E_0$ Coulomb branch at the
blown-up fixed point of the orbifold.
In~\S4.3 we focus on the confinement regime of the quiver theory.
We show that the $SU(3)_1\times SU(3)_2$ flavor symmetry of the quiver is
not broken in this regime; instead, there are nine massless
meson-like composite particles (\ref{CompositeMesons}) which deconstruct the massless
twisted states at the un-blown fixed point.
And in \S4.4 we focus on the transition between the confinement and the Higgs
regimes of the deconstructed theory; in orbifold terms, this transition corresponds
to {\it beginning} to blow up the fixed point.
We show how the slow increase of the scalar VEVs in the quiver theory leads to the abrupt
change of masses of the 4 out of 9 twisted states, from very light to very heavy.
This explain how these 4 states --- which are exactly massless at the on-blown fixed point
and should be light at the early stages of the blow-up --- fail to appear
in the light spectrum of the Ganor--Sonnenschein description of the blown-up
fixed point.

Finally,  section~5 lists the open questions we would like to address in the future.

\section{The Heterotic Model}

The compactification of $E_8 \times E_8$ heterotic string theory on $T^6/\mathbb{Z}_N$
is defined by two vectors. The first of these, the twist vector $\vec{r}$,
defines how the compactified directions are identified under the modding
group. It is three complex-dimensional and depends on the particular $Z_N$
group that the six-torus is modded by. In this work, we are considering
the case $N=3$ with $\vec{r}= (\frac{1}{3},\frac{1}{3},-\frac{2}{3})$. Specifically,
we begin with the complex three-plane, $\mathbb{C}^3$, and mod by the $SU(3)^3$
root lattice, $\Lambda_{SU(3)^3}$, by making the following identifications
for the three complex coordinates $z_i$:
\begin{equation}
z_i \sim z_i+1, \hspace{5mm} z_i \sim z_i + e^{\frac{\pi i}{3}}.
\end{equation}
The result is a specific six-torus, $T^6=\frac{\mathbb{C}^3}{\Lambda_{SU(3)^3}}$.
This $T^6$ is then operated on with a twist parametrized by the twist vector,
identifying points as
\begin{equation}
z_i \rightarrow e^{(2 \pi i) r_i} z_i.
\end{equation}
In all, there are $3^3=27$ fixed points of this twist on the $T^6$ corresponding
to each of the $z_i$'s being either $0$, $\frac{1}{\sqrt{3}}e^\frac{\pi i}{6}$,
or $\frac{2}{\sqrt{3}}e^\frac{\pi i}{6}$.

The other vector, known as the shift vector $\vec{s}$, defines the boundary
conditions of the gauge fields under the twist above. This vector acts like
a Wilson line, breaking the gauge group $E_8 \times E_8$ to some appropriate
subgroup invariant under the twist. Modular invariance highly constrains
the allowed values of $\vec{s}$ so that there are only five distinct consistent
4D $\mathcal{N}=1$ SQFT's differentiated by their resulting gauge groups.
These SQFT's are
\begin{equation}
\begin{aligned}
(0^8;0^8) & : E_8 \times E_8,\\
\frac{1}{3}(1,1,-2,0^5;0^8) & : (E_6 \times SU(3)) \times E_8,\\
\frac{1}{3}(1,1,-2,0^5;1,1,-2,0^5) & : (E_6 \times SU(3)) \times (E_6 \times
SU(3)),\\
\frac{1}{3}(1,1,0^6;-2,0^7) & : (E_7 \times U(1)) \times (SO(14) \times U(1)),\\
\frac{1}{3}(1,1,1,1,-2,0^3;-2,0^7) & : SU(9) \times (SO(14) \times U(1)).\\
\end{aligned}
\end{equation}
Each of these models has their own unique massless spectrum. For each spectrum,
there are untwisted states present everywhere on the orbifold and twisted
states that are located at the fixed points. For our purposes, we are interested
in the theory where $E_8 \times E_8 \to \left( E_6 \times SU(3) \right)
\times \left( E_6 \times SU(3) \right)$, which has untwisted states
\begin{equation}
3 \left( 3,27;1,1 \right) \oplus 3 \left( 1,1;3,27 \right) \oplus \text{9
moduli},
\end{equation}
 and twisted states,
\begin{equation}
27 \left( \overline{3},1; \overline{3},1 \right),
\end{equation}
with one localized at each of the 27 fixed points.

\subsection{The Moduli Space}

Let us consider the twisted states $T^A_{ij}$, $A=1,...,27,\ i,j=1,2,3$ as
27 $3 \times 3$ matrices $T^A$.
The moduli space for these fields must satisfy the F-flatness constraints
stemming from the superpotential
\begin{equation}
W ( T )= \sum_A \det(T^A).
\end{equation}
Specifically, the constraints are
\be
\frac{\partial W}{\partial T^A_{ij}}\ =\ \mathop{\rm minor}(T^A)^{ij}\ =\ 0.
\ee
Note that these constraints do not relate different fixed points to each other.
On the other hand, each $3\times 3$ matrix $T^A$ must have zero minors,
which means that its rank should be at most 1.
Consequently, each $T^A$ must be a tensor product of a row vector and a column vector,
\be
T^A\ =\ u^A\otimes (v^A)^\top,\qquad i.\,e.,\quad
T^A_{ij}\ =\ u^A_i\times v^A_j\,.
\label{Tvev}
\ee

From the $SU(3)_1\times SU(3)_j$ point of view, the $u^A\in({\bf 3},{\bf1})$
while $v^A\in({\bf1},{\bf3})$.
Hence, a non-zero VEV of a $u^A$ triplet breaks $SU(3)_1\to SU(2)_1$ while a
non-zero VEV of a $v^A$ triplet breaks $SU(3)_2\to SU(2)_2$.
However, since the $u^A$ and the $v^A$ are not physical fields but only
their product~(\ref{Tvev}) is physical, the $\vev{T^A}$ VEV leaves an extra
hypercharge $Y=Y_1+Y_2$ unbroken.
Thus, the overall Higgs effect of a single blown-up fixed point is
\be
SU(3)_1\times SU(3)_2\ \to\ SU(2)_1\times SU(2)_2\times U(1)_{1+2}\,.
\label{THiggs}
\ee

Thus far, we have looked at the F-flat directions while ignoring the D-terms due
to Higgsed down $SU(3)_1\times SU(3)_2$.
Unlike the F-terms, the D-terms do relate different fixed points to each other
and require
\be
\sum_A T^A(T^A)^\dagger\ =\,\sum_A (T^A)^\dagger T^A\
=\ \hbox{\Large\bf 1}_{3\times3}\times({\rm a~number}).
\label{Dterms}
\ee
For the $\rm rank=1$ matrices $T^A$, this requires simultaneous blowing up
of at least 3 fixed points with different gauge-group directions of the $u^A$ and $v^A$
triplets, which together break the $SU(3)_1\times U(3)_2$ down to nothing.

However, in this paper we are interested in the individual fixed points
rather than interconnections between them.
Therefore, assume the $T^6$ torus to be very large --- formally, we take the
$\rm radius\to\infty$ limit, --- so that each fixed point may be treated
as an independent $\IC^3/\IZ_3$ singularity.
In 4D field theory terms, the  couplings of the $SU(3)_1\times SU(3)_2$
gauge theories go to zero in the $\rm radius\to\infty$ limit, so the D-terms in the
scalar potential drop to zero, and the constraints (\ref{Dterms}) go away.
Also, in the zero gauge coupling limit, the $SU(3)_1\times SU(3)_2$ can be
thought as a flavor symmetry of each $\IC^3/\IZ_3$ fixed point rather than a gauge symmetry.

\subsection{Geometry}

Instead of compactifying the heterotic string theory and looking at the
massless spectrum, we can attempt to go to the low energy 10D heterotic
supergravity theory first, compactify it on the orbifold, and then look
at the resulting effective 4D theory. This, however, is difficult because
the orbifold is not a manifold and the geometric description at the fixed
points is not well-defined. The procedure, then, would be to blow these
fixed points up, compactify the heterotic supergravity theory on the resulting
Calabi-Yau threefold and then study the blow-down limit.

In general, the fixed point of $\IC^n/\mathbb{Z}_n$ can be smoothed out
by removing the fixed point and replacing it with a $\mathbb{CP}^{n-1}$.
In the limit that this $\mathbb{CP}^{n-1}$ shrinks to zero size, the fixed
point is restored. From the field theory perspective, the resolution of
this fixed point is controlled by the moduli of the theory; giving a VEV
to a scalar field along a flat direction in the moduli space will result
in a deformation of the K\"ahler structure and a smoothing of the corresponding
singularities.

\section{Ho\v{r}ava--Witten Theory}

Similar to type IIA superstring, the
strong string coupling limit of the $E_8 \times E_8$ heterotic string
is dual to the M-theory with a compact eleventh dimension,
but in the heterotic case the $x^{10}$ is compactified on a finite interval
with boundaries \cite{HoravaWitten95,HoravaWitten96}.
The local anomaly cancellation at each boundary requires a 10D $E8$ gauge theory
localized on that boundary, hence the two boundaries of the $x^{10}$ dimension give rise
to the $E8_1\times E8_2$ gauge symmetry of the heterotic string theory.
In the bulk between the two boundaries, the M-theory included the 11D supergravity;
its zero modes on the finite $x^{10}$ interval give rise to the $d=10,\ {\cal N}=1$
supergravity of the heterotic string.

Equipped with this, we see that another possible description of the orbifold
theory would be to first compactify M-theory on the resolved orbifold. The
resulting 5D theory could then be compactified on the interval, where the
singular orbifold limit theory should match that of heterotic supergravity
in the low energy limit. While this is generally not an issue to consider,
there are specific examples of theories that seem to pose a quandary. For
instance, since heterotic M-theory assigns each $E_8$ gauge group to opposite
ends the interval, it is clear to see that states charged under the first
$E_8$ reside on one end, while states charged under the other $E_8$ will
be on the opposite end. However, there are unique cases, such as the one
we are studying, where a state is somehow charged under both $E_8$'s. The
bulk theory that separates these two boundaries is simply 11D supergravity,
so it is highly nontrivial to describe how a state could be charged across
this bulk.

In fact, such a description was found for the somewhat simpler case of heterotic
M-theory compactified on a Calabi-Yau twofold to a 6D field theory \cite{Vadim99,Vadim01}.
It was shown that the theories at these fixed points could be described
as a 7D SYM theory compactified on the interval with particular boundary
conditions. It was also shown that the fixed points could be described locally
as multi--Taub--NUT spaces. This allowed for the alternative description in
terms of type I' string theory, and brane-engineering was employed to naturally
describe the theory and boundary conditions.

Unfortunately, the generalization of this procedure from 6D to 4D becomes
complicated. First, there is no 6D equivalent to the multi-Taub-NUT space
at the fixed points of the orbifold, so the procedure using brane engineering
is not valid. Second, the 5D theories present at the fixed points when M-theory
is compactified on an orbifold are not SYM theories as we would expect from
the case above, but 5D interacting superconformal field theories \cite{MorrisonSeiberg96}.
Compactifying these theories on an interval is extremely nontrivial, as
the theories themselves can be quite difficult to study.

\subsection{The Bulk Theory: $E_0$ SCFT}

Let's now discuss M-theory compactified on $T^6/\mathbb{Z}_3$ in more detail.
As stated above, the orbifold can be smoothed out to a Calabi-Yau threefold
by replacing the 27 singular fixed points with $\mathbb{CP}^2$'s. Blowing
these $\mathbb{CP}^2$'s down to points will, in turn, reproduce the orbifold.
More generally, $\mathbb{CP}^2$ is a del Pezzo surface, and it is has been
shown \cite{MorrisonSeiberg96} that M-theory compactified on a Calabi-Yau
threefold with a collapsing del Pezzo surface results in a 5D $\mathcal{N}=1$
SCFT at that point with a global $E_N$ symmetry. In the case of $\mathbb{CP}^2$,
the resulting SCFT has $E_0$ global symmetry, which is no global symmetry
at all.

The $E_0$ SCFT can best be illustrated in terms of brane webs \cite{AHK97,
KolRahmfeld98}. The webs are formed by $(p, q)$ 5-branes in Type IIB string
theory.  They share $(x^0,x^1,x^2,x^3,x^4)$, and form $(p, q)$-lines in
the $(x^5,x^6)$-plane (for an appropriate choice of complexified string
coupling $\tau=\chi+ie^{-\phi_s}$, namely $\tau=i$). Different $(p, q)$-lines
are allowed to meet as long as $(p, q)$-charge is conserved:
\begin{equation}
\sum p_i=\sum q_i=0.
\end{equation}
The resulting webs formed by these $(p, q)$ 5-branes in the $(x^5,x^6)$-plane
describe 5D $\mathcal{N}=1$ $SU(N)$ gauge theories in the common $(x^0,x^1,x^2,x^3,x^4)$
volume.

\begin{figure}[H]
\centering
\begin{pspicture}(-0.5,-2.25)(15,+2.5)
\rput(3,0){
	\psline[linestyle=dashed](-3,-2)(-1,0)(-3,+2)
	\psline[linestyle=dashed](+3,-2)(+1,0)(+3,+2)
	\psline[linestyle=dashed](-1,0)(+1,0)
	\psframe(-1.7,-0.7)(+1.7,+0.7)
	\psline(-3,-2)(-1.7,-0.7)
	\psline(-3,+2)(-1.7,+0.7)
	\psline(+3,-2)(+1.7,-0.7)
	\psline(+3,+2)(+1.7,+0.7)
	\rput[t](0,+2.25){$\theta=0$}
	}
\rput(10,0){
	\psline[linestyle=dashed](-3,-2)(-1,0)(-3,+2)
	\psline[linestyle=dashed](+1,-2)(+1,0)(+5,+2)
	\psline[linestyle=dashed](-1,0)(+1,0)
	\pspolygon(-1.7,-0.7)(-1.7,+0.7)(+2.4,+0.7)(+1,-0.7)
	\psline(-3,-2)(-1.7,-0.7)
	\psline(-3,+2)(-1.7,+0.7)
	\psline(+1,-2)(+1,-0.7)
	\psline(+5,+2)(+2.4,+0.7)
	\rput[t](0,+2.25){$\theta=\pi$}
	}
\end{pspicture}
\caption{$SU(2)$ SYM with $\theta=0$ and $\theta=\pi$.}
\label{both_thetas}
\end{figure}
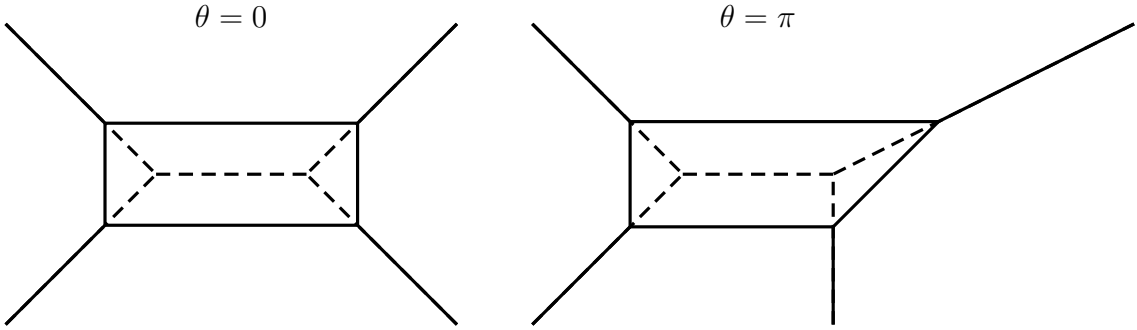

For our interests, let us consider $SU(2)$ supersymmetric Yang-Mills theory.
Similar to 4D where $\pi_3 (SU(2))=\mathbb{Z}$ leads to a vacuum $\theta$-angle
that can take values in \{$2\pi \mathbb{Z}$\}, in 5D we have $\pi_4 (SU(2))=\mathbb{Z}_2$
which leads to a $\theta$-angle that can take values \{$0, \pi$\}. We thus
have two $SU(2)$ SYM theories, differing by the value of a $\theta$-angle.
The corresponding brane webs take the forms in Fig. \ref{both_thetas}. In
both cases, the VEV of the real scalar field $\phi$ in the vector multiplet
is associated to the ``breathing" mode, or contraction and expansion, of
the quadrilateral. As can be seen, varying $\phi$ does not affect the asymptotic
configuration of the external legs in the $(x^5,x^6)$-plane, and hence parametrizes
a local symmetry, i.e., the gauge symmetry. When $\phi=0$, the boxes collapse,
as is illustrated by the dotted lines. In this limit, the length of the
horizontal dotted line in each corresponds to its bare coupling $h=4\pi^2/g^2_0$.
Varying $h$ changes the asymptotic configuration, signaling that it parametrizes
a global symmetry. Specifically, it corresponds to a global $U(1)$ symmetry
associated to the conserved instanton current $j=\ast (F \wedge F)$ that
can be defined in 5D.

\begin{figure}[ht]
\centering
\psset{unit=0.043\hsize,linewidth=1pt}
\begin{pspicture}(0,-3.5)(23.25,+9.5)
\rput(3,7){%
    \psline[linestyle=dashed](-3,-2)(-1,0)(-3,+2)
    \psline[linestyle=dashed](+1,-2)(+1,0)(+5,+2)
    \psline[linestyle=dashed](-1,0)(+1,0)
    \pspolygon(-1.75,-0.75)(-1.75,+0.75)(+2.5,+0.75)(+1,-0.75)
    \psline(-1.75,-0.7)(-3,-2)
    \psline(-1.75,+0.7)(-3,+2)
    \psline(+2.5,+0.7)(+5,+2)
    \psline(+1,-0.7)(+1,-2)
    \rput[t](0.5,2.5){(a)}
	}
\rput(11.5,7){%
    \psline[linestyle=dashed](-2.2,-2.2)(0,0)(-2.2,+2.2)
    \psline[linestyle=dashed](0,-2.2)(0,0)(+4.4,+2.2)
    \pspolygon(-0.9,-0.9)(-0.9,+0.9)(+1.8,+0.9)(0,-0.9)
    \psline(-0.9,-0.9)(-2.2,-2.2)
    \psline(-0.9,+0.9)(-2.2,+2.2)
    \psline(+1.8,+0.9)(+4.4,+2.2)
    \psline(0,-0.9)(0,-2.2)
    \rput[t](0.5,2.5){(b)}
	}
\rput(18.25,7){%
    \psline[linestyle=dashed](0,+1.2)(+0.8,+0.4)(+2.4,+1.2)
    \psline[linestyle=dashed](+0.8,+0.4)(0,-1.2)
    \pspolygon(0,-1.2)(0,+1.2)(+2.4,+1.2)
    \psline(0,-1.2)(-1.3,-2.5)
    \psline(0,+1.2)(-1.3,+2.5)
    \psline(+2.4,+1.2)(+5,+2.5)
    \psline(0,-1.2)(0,-2.5)
    \rput[t](1,2.5){(c)}
	}
\rput(1.8,0){%
    \psline[linestyle=dashed](0,+1.2)(+0.8,+0.4)(+2.4,+1.2)
    \psline[linestyle=dashed](+0.8,+0.4)(0,-1.2)
    \pspolygon(0,-1.2)(0,+1.2)(+2.4,+1.2)
    \psline(0,+1.2)(-1.3,+2.5)
    \psline(+2.4,+1.2)(+5,+2.5)
    \psline(0,-1.2)(-0.5,-2.2)
    \psline(-0.5,-3.5)(-0.5,-2.2)(-1.8,-3.5)
    \rput[t](1,2.75){(d)}
    }
\rput(9.5,0){%
    \psline[linestyle=dashed](0,+1.2)(+0.8,+0.4)(+2.4,+1.2)
    \psline[linestyle=dashed](+0.8,+0.4)(0,-1.2)
    \pspolygon(0,-1.2)(0,+1.2)(+2.4,+1.2)
    \psline(0,+1.2)(-1.3,+2.5)
    \psline(+2.4,+1.2)(+5,+2.5)
    \psline(0,-1.2)(-0.9,-3)
    \rput[t](1.25,2.75){(e)}
	}
\rput(17.2,0){%
    \pscircle*(+0.8,+0.4){0.2}
    \psline(+0.8,+0.4)(-1.3,+2.5)
    \psline(+0.8,+0.4)(+5,+2.5)
    \psline(+0.8,+0.4)(-0.9,-3)
    \rput[t](1.5,2.75){(f)}
	}
\end{pspicture}
\caption{%
Brane webs for the $SU(2)_{\theta=\pi}$ SYM and the $E_0$ SCFT.
$\phi>0$ for all 6 webs, while $h$ keeps decreasing:
(a) $h>0$;
(b) $h=0$;
(c) $h=h_{\rm flop}=-\phi<0$;
(d) $-3\phi<h<-\phi$;
(e) $h\to-\infty$, $\phi\to+\infty$ but finite $\hat\phi=\phi+\tfrac13 h>0$
--- the $E_0$ Coulomb branch;
(f) $h\to-\infty$, $\phi\to+\infty$, $\hat\phi=0$ --- the superconformal $E_0$.
}
\label{fig:ManyBraneWebs}
\end{figure}
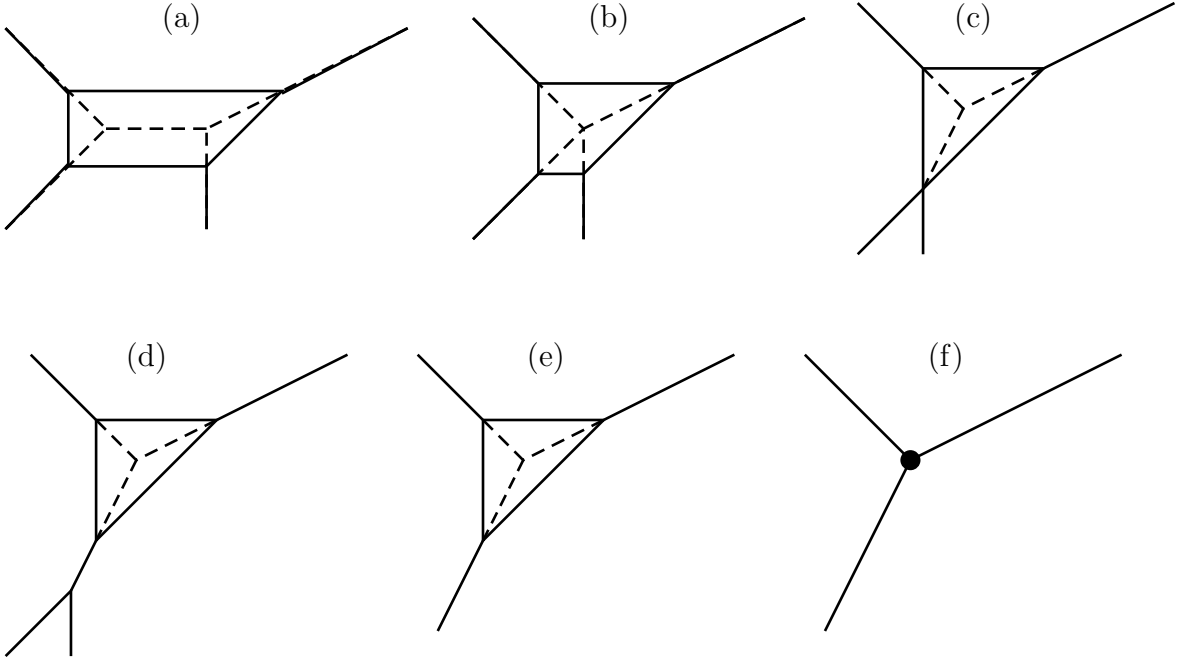

To investigate the $E_0$ SCFT, we will need to start with $SU(2)$
SYM with $\theta=\pi$.
We can vary the parameters $\phi$ and $h$, and
explore the various limits that result.
Starting at $h>0$, $\phi>0$ as
in fig.~\ref{fig:ManyBraneWebs}.(a), we can let $h \to 0$.
At $h=0$, the quantum corrected
coupling is still positive, so the theory is still $SU(2)$ SYM with $\theta=\pi$,
as in fig.~\ref{fig:ManyBraneWebs}.(b).
As we continue into negative $h$, we eventually
reach a point $h=h_{\rm flop}$ where the coupling diverges and a quark becomes
massless (the mass depends on the length of the bottom brane, which goes
to zero at $h_{\rm flop}$, fig.~\ref{fig:ManyBraneWebs}.(c)).
Continuing past this point, there
is a flop transition as seen in fig.~\ref{fig:ManyBraneWebs}.(d).
This new phase has a massive quark with a mass proportional to $h$.
In the low energy effective theory,
we only care about the massless spectrum, so we discard this massive quark.
The resulting theory --- represented by the brane web in fig.~\ref{fig:ManyBraneWebs}.(e)
--- is the $E_0$ theory along its Coulomb branch.
Its massless spectrum consists
of a single $U(1)$ vector multiplet whose scalar field $\hat{\phi}$, a linear
combination of $\phi$ and $h$, characterizes the breathing mode of the
resulting triangle.
There is no possible global deformation, so there is
no global symmetry in the $E_0$ theory, as stated earlier.
When $\hat{\phi}=0$
as in Fig.\ \ref{fig:ManyBraneWebs}.(f), the coupling diverges and we reach the $E_0$ SCFT
point in the moduli space.

\subsection{The Boundaries of $E_0$}

Now that we have established M-theory compactified on $T^6/\mathbb{Z}_3$,
let's compactify it further on $S^1/\mathbb{Z}_2$ to relate it to $E_8 \times
E_8$ heterotic string theory. At the fixed point, this amounts to compactifying
the $E_0$ SCFT on $S^1/\mathbb{Z}_2$. This introduces boundaries and, as
a result, anomalies. Specifically, the 11D supergravity Chern-Simons term
introduces anomalies. Ganor and Sonnenschein investigated this in \cite{GanorSon02}
and found that, when compactified to 5D, the resulting Chern-Simons term
introduces three times the usual anomaly to each boundary. Looking at the
low-energy effective theory of $E_0$ along the Coulomb branch with $\hat{\left<\phi
\right>} \gg 1/R$, they canceled the anomalies at each boundary with the
addition of 3 4D chiral multiplets $X$ with $U(1)$ charge $+1$ at $x^4=0$
and 3 4D chiral multiplets $Y$ with $U(1)$ charge $-1$ at $x^4=\pi$. The
$X$'s transform under the fundamental of an $SU(3)_L$ global symmetry, and
the $Y$'s transform under the fundamental of another $SU(3)_R$ global symmetry.
The introduction of these chiral multiplets modifies the D-term, which imposes
the boundary conditions
\begin{equation}
\frac{1}{2}\hat{\phi}^2|_{x^4=0}
= \left| X_0 \right| ^2, \hspace{1in} \frac{1}{2}\hat{\phi}^2|_{x^4=\pi}=
\left| Y_0 \right| ^2.
\end{equation}
where $X_0$, $Y_0$ are the scalars in the corresponding multiplets. We can
thus see that the VEVs of these boundary fields are related to the $E_0$
scalar field VEV.

To make contact with the theory at the $T^6/\IZ_3$ fixed points, it is necessary
to extrapolate from these results down to $\hat{\left<\phi \right>} \ll
1/R$, specifically $\hat{\left<\phi \right>}=0$. Ganor and Sonnenschein
established a conjecture in \cite{GanorSon02} that the global $SU(3)_L \times
SU(3)_R$ is generically broken to $SU(2)_L \times SU(2)_R \times U(1)_V$
for $\hat{\left<\phi \right>} \neq 0$. In this case, the boundary fields
introduced for anomaly cancellation have charges
\begin{equation}
X:(2,1)_{(1,1)}+(1,1)_{(-2,1)}, \hspace{.75in} Y:(1,2)_{(-1,-1)}+(1,1)_{(2,-1)},
\end{equation}
in notation $(SU(2)_L,SU(2)_R)_{(U(1)_V,U(1)_B)}$, where $U(1)_B$ is the
gauge symmetry. The $X$ and $Y$ VEVs are not invariant under $U(1)_V$,
but are invariant under a combination of $U(1)_V$ and $U(1)_B$, with charge
$Q_C\equiv \frac{1}{2}Q_V+Q_B$. Rewriting the charges now as $(SU(2)_L,SU(2)_R)_{U(1)_C}$,
we have
\begin{equation} \label{state_charges}
X:(2,1)_{\frac{3}{2}}+(1,1)_{0}, \hspace{.75in} Y:(1,2)_{-\frac{3}{2}}+(1,1)_{0}.
\end{equation}
\indent Let us examine the anomalies due to these states at one of the boundaries,
say $x^4=0$. If there is to be $SU(3)_L \times SU(3)_R$ restoration at $\hat{\left<\phi
\right>}=0$, then 't Hooft anomaly matching requires that the $SU(3)_L^3$
triangle anomaly must match the $SU(2)_L^2-U(1)_C$ triangle anomaly. In
order for this to occur in the presence of $X$, it is necessary for there
to be three triplets under $SU(3)_L$. Similar analysis at the $x^4=\pi$
boundary indicates that we need three triplets under $SU(3)_R$. The simplest
collection of states that satisfies this requirement is a single state with
charge $(3,3)$ under $SU(3)_L \times SU(3)_R$. This is at least suggestive
that a state of this form could be present at the restoration point $\hat{\left<\phi
\right>}=0$.

\section{Dimensional Deconstruction}
In this section, we  deconstruct the fifth dimension $x^4$ (n\'ee $x^{10}$)
of the $E_0$ SCFT living on a fixed plane of the $T^6/\IZ_3$ orbifold.
But let us start with a brief review of the  {\it dimensional deconstruction}
\cite{Arkani01} in general.
Basically, it amounts to latticizing one or more dimensions of a theory 
followed by  reinterpreting  the lattice as a quiver.
For example, consider a 5D Yang--Mills theory with an $SU(N)$ gauge group.
We begin by discretizing the fifth dimension with some lattice spacing $a$
while keeping the other dimensions $x^{0,1,2,3}$ continuous.
On the resulting lattice, the $A^{0,1,2,3}$ components
of the vector field $A^M(x)$ live on the nodes $x^4=\ell\times a$ while the $A^4$
lives on the lattice links, encoded in unitary matrices
\be
U_\ell\ =\ \mathop{\rm Pexp}\left( \int_{\ell a}^{(\ell+1)a} A_4\,dx^4 \right) .
\ee
Next, we reinterpret the $A^{0,1,2,3}(x^{0,1,2,3},\ell)$ as 4D gauge fields
of the $\ell^{\rm th}$ factor of
\be
G_{\rm 4D}\ =\ \prod_\ell SU(N)_\ell\,,
\label{ProdSUN}
\ee
and the $U_\ell(x^{0,1,2,3})$ as 4D scalar fields in bifundamental multiplets
$(\square_\ell,\overline\square_{\ell+1})$ of the gauge group~(\ref{ProdSUN}),
hence the whole theory can be described by the quiver
\be
\begin{pspicture}(0,-0.3)(12,+0.3)
\psline[linestyle=dotted,linecolor=orange,linewidth=3pt](0,0)(12,0)
\multido{\n=1+2}{6}{%
	\pscircle*[linecolor=green](\n,0){0.4}
	\rput(\n,0){$\scriptstyle N$}
	}
\end{pspicture}
\ee
To make the 4D quiver theory renormalizable, we may replace the non-linear
scale fields $U_\ell$ with the linear bifundamental fields subject to a scalar potential
with degenerate minima spanning the $SU(N)$ group manifold.
Alternatively, we may realize the $U_\ell$ as technipions of confining gauge theories
$SU(M)_\ell$ with $N$ massless flavors.

For another example, consider the  $\rm SQCD_{5D}$ with $n_c$ colors and $n_f$ flavors.
As explained in \cite{Vadim04,Vadim06}, this theory deconstructs to the 4D quiver
\be
\psset{unit=4mm,arrowscale=1.5}
\def\site{%
    \psset{linewidth=1pt,linecolor=red,arrowscale=1.2}
    \pscircle*[linecolor=green](0,0){1}\relax
	\rput{*0}(0,0){$n_c$}\relax
	\psline{->}(-1.8,+0.45)(-0.9,+0.45)\relax
    \psline{->}(-1.8,-0.45)(-0.9,-0.45)\relax
    \psline{->}(-1.8,-0.15)(-0.99,-0.15)\relax
    \psline{->}(-1.8,+0.15)(-0.99,+0.15)\relax
    \psline{<-}(+1.8,+0.45)(+0.9,+0.45)\relax
    \psline{<-}(+1.8,-0.45)(+0.9,-0.45)\relax
    \psline{<-}(+1.8,-0.15)(+0.99,-0.15)\relax
    \psline{<-}(+1.8,+0.15)(+0.99,+0.15)\relax
    }
\begin{pspicture}(1,-2)(39.5,+2)
\multido{\n=5+5}{7}{\rput(\n,0){%
	\rput{90}{\site}
	\psline[linecolor=blue]{->}(1,0)(3,0)
	\psline[linecolor=blue](-1,0)(-2.5,0)
	}}
\psline[linecolor=blue,linewidth=1.5pt,linestyle=dotted]{->}(1,0)(3,0)
\psline[linecolor=blue,linewidth=1.5pt,linestyle=dotted](37.5,0)(39.5,0)
\end{pspicture}
\label{SQCD5}
\ee
The deconstruction preserves 4 out of 8 supercharges of the 5D SUSY, so the quiver has
${\cal N}=1$ SUSY in 4D.
Together, the 4D vector multiplets and the bifundamental chiral multiplets of the quiver
deconstruct the 5D vector multiplets.
Some of the 4D anti/fundamental fields deconstruct the 5D quarks and antiquarks,
while additional 4D anti-fundamentals may be used to adjust the Chern--Simons level
of the 5D theory (or the $\theta$ angle for $n_c=2$).

\subsection{Deconstructing $E_0$ Without Boundaries}
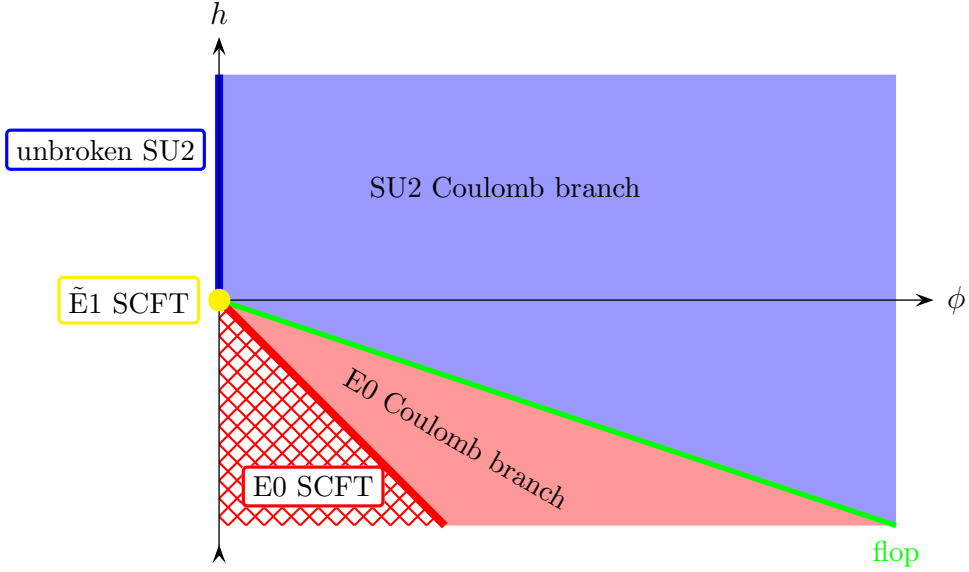
\begin{figure}[ht]
\centering
\psset{unit=10mm,framearc=0.2}
\begin{pspicture}(-2,-3.5)(10,+3.7)
\pspolygon*[linecolor=paleblue](0,0)(0,+3)(9,+3)(9,-3)
\pspolygon*[linecolor=palepink](0,0)(9,-3)(3,-3)
\pspolygon[linestyle=none,fillstyle=crosshatch,hatchcolor=red](0,0)(0,-3)(+3,-3)
\rput[l](2,1.5){\small SU2 Coulomb branch}
\rput[l]{330}(1.7,-1){\small E0 Coulomb branch}
\psline[linecolor=blue,linewidth=3pt](0,0)(0,+3)
\uput[l](0,+2){\psframebox[linecolor=blue]{\small unbroken SU2}}
\psline[linewidth=2pt,linecolor=green](0,0)(9,-3)
\rput[t](9,-3.2){\green\small flop}
\rput[tr](2.2,-2.2){\psframebox*[framesep=0pt]{\psframebox[linecolor=red]{\small E0 SCFT}}}
\psline[linewidth=3pt,linecolor=red](0,0)(3,-3)
\psline[linewidth=0.5pt,arrowscale=2]{>->}(0,-3.5)(0,+3.5)
\uput[u](0,+3.5){$h$}
\psline[linewidth=0.5pt,arrowscale=2]{->}(0,0)(9.5,0)
\uput[r](9.5,0){$\phi$}
\pscircle*[linecolor=yellow](0,0){0.15}
\rput*[r](-0.25,0){\psframebox[linecolor=yellow]{\small \~E1 SCFT}}
\end{pspicture}
\caption[]{%
    Moduli/parameter space of the 5D $SU(2)_{\theta=\pi}$ SYM,
    the $E_0$ SCFT, and their Coulomb branches.
    The horizontal axis is the Coulomb modulus $\phi$ of the $SU(2)$
    while the vertical axis is the inverse gauge coupling $h$ according to\\
    \leavevmode
    \begin{minipage}{\linewidth}
    \be
    h\,=\, {4\pi^2\over g^2_{\rm 5d}[{\rm SU2}]}\,,\qquad
	\vev{\Phi_{SU2}}\,=\begin{pmatrix} +\phi & 0\\ 0 & -\phi\\ \end{pmatrix}.
	\label{SU2moduli}
	\ee
	\end{minipage}
	The $E_0$ Coulomb modulus is the $\hat\phi=\phi+\frac13 h$.
	}
\label{ModuliSpaceDiagram}
\end{figure}

Now let's turn our attention to the $E_0$ theory in  infinite $4+1$ dimensions;
we shall deal with the finite $x^4$ (n\'ee $x^{10}$) dimension in the next subsection.
There is no standard procedure for deconstructing superconformal theories,
so we are going to exploit the connection between the moduli/parameter spaces of the
$E_0$ SCFT and the $SU(2)$ SYM with $\theta=\pi$,
see figure~\ref{ModuliSpaceDiagram} for the phase diagram and figures
\ref{fig:ManyBraneWebs}.(a--f) for the brane webs for all the phases.
Thus, we shall start by deconstructing the $SU(2)_{\theta=\pi}$ theory, go to the
Coulomb branch, identify the flop transition to the $E_0$ Coulomb branch,
and eventually reach the SCFT point in the moduli space.

As explained in \cite{Vadim04}, the 5D $SU(2)$ SYM with $\theta=\pi$ deconstructs
to the following 4D quiver
\be
\psset{unit=10mm,linewidth=1.2pt,arrowscale=1.5}
\def\node(#1){%
	\rput(#1){%
		\pscircle*[linecolor=green](0,0){0.4}
		\rput(0,0){2}
		\psline[linecolor=red]{<-}(0,+0.4)(0,+1.2)
		\psline[linecolor=red]{->}(0,-0.4)(0,-1.2)
		\psline[linecolor=blue]{->}(0.4,0)(1.6,0)
		\uput[r](0,+0.9){$q$}
		\uput[r](0,-0.9){$\tilde q$}
		\uput[u](1,0){$\Omega$}
		}
	}
\begin{pspicture}(-2,-1.5)(12,+1)
\multido{\n=0+2}{6}{\node(\n,0)}
\psline[linecolor=blue,linestyle=dotted](-2,0)(-1.6,0)
\psline[linecolor=blue]{->}(-1.6,0)(-0.4,0)
\uput[u](-1,0){$\Phi$}
\psline[linecolor=blue,linestyle=dotted](11.6,0)(12,0)
\end{pspicture}
\label{SU2quiver}
\ee
with the superpotential
\be
W\ =\,\sum_\ell s_\ell\bigl(\det(\Phi_\ell)\,-\,v^2\bigr)\
+\ \sum_\ell \tilde q_\ell\,\Omega_\ell\, q_{\ell+1}
\label{SU2quiverW}
\ee
where the $s_\ell$ are gauge singlets (not shown on the quiver~(\ref{SU2quiver})).
Each $SU(2)_\ell$ factor of the quiver has the same 4D gauge coupling $g$, or in quantum terms
the same $\Lambda_\ell=\Lambda$; the 5D gauge coupling of the deconstructed $SU(2)_{\rm 5D}$ obtains
as
\be
h\ =\ {4\pi^2\over g_5^2}\
=\ {1\over a}\,\log\left|{v\over\Lambda}\right|^3
\label{5Dcoupling}
\ee
where $a\approx1/(gv)$ is the lattice spacing.
Finally, the moduli space of the quiver is the complexified Coulomb moduli space
of the $SU(2)_{\rm 5D}$; it's parametrized by equal (modulo gauge symmetries)
VEVs of all the bifundamental scalars,
\be
{\rm all}\quad \vev{\Omega_\ell}\
=\,\begin{pmatrix} \omega_+ & 0 \\ 0 & \omega_- \\ \end{pmatrix},\quad
\omega_\pm\ =\ v\times\exp\bigl( \pm a\phi\bigr)
\label{OmegaVEVs}
\ee
where $\Re\phi$ is the 5D modulus from eq.~(\ref{SU2moduli})
while $\Im\phi$ is irrelevant for an infinitely long quiver;
without loss of generality we shall henceforth assume real $\phi>0$.
The $\phi=0$ point --- and hence
$$
{\rm all}\quad \vev{\Omega_\ell}\
=\,\begin{pmatrix} v & 0 \\ 0 & v \\ \end{pmatrix}
$$
--- corresponds to the unbroken $SU(2)$ in 5D while $\phi>0$ spans the Coulomb branch.

When $|\Lambda|\ge|v|$, the quiver theory becomes strongly coupled in the IR.
For the deconstructed theory, this means $h<0$, which puts us in the bottom half
of the moduli/paramater space~(\ref{ModuliSpaceDiagram}).
For $h=-\phi$ in that bottom half, we should have a flop transition to the $E_0$
Coulomb branch.
To see how this works in the quiver, consider the mass terms for the fundamental
$Q_\ell$ and $\tilde Q_\ell$ fields as functions of the $\omega_\pm$:
\be
W\
\supset\ \omega_+ \times \sum_\ell \tilde q_{\ell,(1)} q_{\ell+1}^{(1)}\
+\ \omega_- \times \sum_\ell \tilde q_{\ell,(2)} q_{\ell+1}^{(2)}\
+\ {\Lambda^3\over\omega_+^2}\times\sum_\ell \tilde q_{\ell-1,(2)} q_{\ell+1}^{(2)}\,,
\label{QuarkMasses}
\ee
where the first two terms on the RHS are tree-level while the third term
stems from the one-instanton effects in the the $SU(2)_\ell$ Higgsed down by the
$\vev{\Omega_{\ell-1}}$ and $\vev{\Omega_{\ell}}$.\footnote{%
	To see the origin of these instanton terms, let's focus on a single $SU(2)_\ell$
	gauge group factor and ignore all the others.
	Let's temporarily turn on VEVs of the $\tilde q_{\ell-1,(2)}$
	and $q_{\ell+1}^{(2)}$ scalars (which are both neutral WRT $SU(2)_\ell$)
	while turning off the $\omega_-$ eigenvalue of
	the bifundamental VEVs $\vev{\Omega_{\ell-1}}$ and $\vev{\Omega_\ell}$.
	The $\vev{\tilde q_{\ell-1,(2)}}$ VEV gives mass to the $SU(2)_\ell$ doublets
	$q_\ell^{(\alpha)}$ and $\Omega_{\ell-1,(\alpha)}^{(2)}$ ($\alpha=1,2$), while the
	$\vev{q_{\ell+1}^{(2)}}$ VEV gives masses to the $\tilde q_{\ell,(\alpha)}$
	and $\Omega^{(\alpha)}_{\ell,(2)}$ doublets.
	Integrating out these doublets from the $SU(2)_\ell$ gauge theory leaves us
	with two massless doublets $\Omega_{\ell-1,(\alpha)}^{(1)}$
	and $\Omega^{(\alpha)}_{\ell,(1)}$ and effective strong-interaction scale
	$\Lambda^5_{\rm eff}=\Lambda^3\vev{\tilde q_{\ell-1,(2)}}\vev{q_{\ell+1}^{(2)}}$.
	The remaining doublets have $\omega_+$ VEVs which Higgs the $SU(2)_\ell$ down
	to nothing exactly as in the Affleck--Dine--Seiberg setup \cite{ADS},
	and just like in that setup, the instantons of the broken gauge theory generate
	the superpotential
	\be
	W_{\rm inst}\ =\ {\Lambda^5_{\rm eff}\over\omega_+^2}\
	=\ {\Lambda^3\vev{\tilde q_{\ell-1,(2)}}\vev{q_{\ell+1}^{(2)}}\over\omega_2}
	\label{Winstanton}
	\ee
	for the $\Omega_{\ell-1,(\alpha)}^{(1)}$ and $\Omega^{(\alpha)}_{\ell,(1)}$
	doublets.
	Now let's analytically continue this one-instanton superpotential to
	zero $\vev{\tilde q_{\ell-1}}$ and $\vev{q_{\ell+1}}$ (but non-zero $\omega_+$).
	In this regime, the superpotential (\ref{Winstanton}) becomes the
	$\omega_+$ dependent mass term for the  $\tilde q_{\ell-1,(2)}$
	and $q_{\ell+1}^{(2)}$ fields,
	\be
	W_{\rm inst}\ =\ {\Lambda^3\over\omega_+^2}\times
	\tilde q_{\ell-1,(2)}\,q_{\ell+1}^{(2)}\,.
	\ee
	Adding such mass terms for all the $SU(2)_\ell$ gauge groups of the quiver
	gives us the double-hopping third term in the superpotential (\ref{QuarkMasses}).
	}
The eigenvalues of the doublets' mass matrix follow from the Fourier transform
from $\ell$ to the momentum $p_4$ in the $x^4$ direction, thus
for the $SU(2)$ color $\alpha=1$
\be
m_1(p_4)\ =\ \omega_+\times e^{iap_4},
\ee
while for the $\alpha=2$ color
\be
m_2(p_4)\ =\ \omega_-\times e^{iap_4}\ +\ {\Lambda^3\over\omega_+^2}\times e^{2iap_4} .
\label{QM2}
\ee
For generic values of the eigenvalues $\omega_\pm$ these masses never come close
to zero,  so the effective low-energy theory is simply
the deconstructed $U(1)_{\rm 5D}\subset SU(2)_{\rm 5D}$ of the Coulomb branch.
However, when
\be
\left|\omega_-\right|\ =\ \left|{\Lambda^3\over\omega_+^2}\right|,
\label{Qflop}
\ee
the mass $m_2(p_4)$ crosses zero for some momentum $p_4$.
Without loss of generality we may assume this happens for $p_4=0$
(otherwise, we can shift the $p_4$ by a constant by a suitable redefinition of
the quark field phases), thus $m_2\sim p_4+\rm lattice$ corrections, which means
the deconstructed 5D quark with color $\alpha=2$ has zero 5D mass.
And the massless charged particle is exactly what should happen at the flop transition!
Moreover, the flop condition~(\ref{Qflop}) corresponds in 5D terms to
\be
v\times e^{-a\phi}\ =\ v\times e^{-2a\phi}\times e^{-ah}\quad
\Longleftrightarrow\quad h\ =\ -\phi,
\ee
which is precisely where the 5D $SU2_{\theta=\pi}$ SYM should have the flop
transition to the $E_0$ Coulomb branch.

Going further below the flop transition line of the diagram (\ref{ModuliSpaceDiagram})
we eventually reach the line of the superconformal $E_0$ at
$h=-3\phi$.
On this line, the 5D moduli space ends --- the $E_0$ Coulomb modulus
\be
\hat\phi\ =\ \phi\ +\ \tfrac13 h\
\ee
never becomes negative in infinite $4+1$ dimensions.
Although the $E_0$ compactified on a circle allows for $\hat\phi<0$,
the negative--$\hat\phi$ chamber of the moduli space shrinks to nothing
in the decompactification limit due to $g_{\hat\phi\hat\phi}\to 0$.

In the quiver terms, the $E_0$ modulus is
\be
\hat\phi\ =\ {1\over a}\,\log\left|{\omega_+\over\Lambda}\right|
\label{E0modulus}
\ee
while the superconformal line $\hat\phi=0$ (for $\phi>0$ and $h<0$) corresponds to
\be
|\omega_+|\ =\ |\Lambda|\ \gg\ |v|\ \gg |\omega_-|
\label{SCline}
\ee
From the 4D point of view, this line is the transitions between the semiclassical
Higgs regime of the quiver theory and the confinement regime.
Indeed, for $\omega_+\gg\Lambda$, we have semiclassical Higgsing of the $\prod_\ell SU(2)_\ell$
theory down to a single $U(1)$, with a Kaluza-Klein tower of light 4D photons
corresponding to the deconstructed $U(1)_{\rm 5D}$.
On the other hand, for $|\Lambda|\gg|\omega_+|\ge|\omega_-|$,
the Higgs effects of the scalar VEVS become negligible compared to the non-perturbative
4D effects such as confinement.

For a single $SU(2)$ gauge theory, there is a smooth crossover between the Higgs
and the confinement regimes of the theory rather than a phase transition.
But for for the $\bigl[SU(2)\bigr]^N$ quiver theory with $N\to\infty$, the transition
seems to become abrupt.
Thus, any scalar $\rm VEV>\Lambda$ --- even if its just a little bit larger than $\Lambda$ ---
puts the quiver in the semiclassical Higgs regime.
On the other hand, a scalar $\rm VEV<\Lambda$ is as good as zero.
So when all the VEVs are smaller than $\Lambda$ --- even if they are just a hair smaller ---
the quiver is in the confinement regime, and the Higgs effects of the scalar VEVS are
of no importance.

For the quivers with large but finite $N$, we expect to have a continuous crossover between
the Higgs and the confinement regimes, but the crossover should become sharper and sharper
with larger $N$.
As a heuristic explanation of this behavior, note that the  holomorphic gauge-invariant
order parameters of the quiver behave line $({\rm VEV}/\Lambda)^N$, so the crossover
between the large-VEV and small-VEV regimes should become sharper and sharper with
larger $N$.
Ultimately, in the $N\to\infty$ limit, the crossover becomes infinitely sharp and turns
into an abrupt phase transition.

For the quiver (\ref{SU2quiver}) at hand, this means that
\begin{itemize}
\item[$\star$]
For $\red\Lambda\ge\omega_+$, {\it every $SU(2)_\ell$ of the quiver confines}
and the effect of the scalar eigenvalues $\omega_\pm$ is negligible.
This regime deconstructs the $E_0$ SCFT.
\item[$\star$]
For $\red\Lambda<\omega_+$, {\it every $SU(2)_\ell$ is Higgsed down}
and only the diagonal $U(1)_{\rm diag}\subset SU(2)_{\rm diag}\subset [SU(2)]^N$
survives.
This regime deconstructs the Coulomb branch of the $E_0$ (or perhaps the Coulomb branch
of the $SU(2)_{\rm \theta=\pi}$).
\end{itemize}

This completes our survey of the deconstructed $SU(2)+E_0$ moduli space.

Since in this paper we are interested in the $E_0$ theory rather than the $SU(2)$ SYM,
we are going to take the limit of $\phi\to+\infty$, $h\to-\infty$ while $\hat\phi$
stays finite, see figures~\ref{fig:ManyBraneWebs}.(d--e) for the brane web illustration.
In quiver terms, this corresponds to setting $v=0$ and hence enforcing $\omega_-=0$
while $\omega_+$ remains unconstrained, thus
\be
\vev{\Omega_\ell}\ =\,\begin{pmatrix} \omega_+ & 0 \\ 0 & 0 \\ \end{pmatrix},\quad
\hat\phi\ =\ {1\over a}\,\log\left|{\omega_+\over\Lambda}\right| .
\label{E0vev}
\ee
Thanks to the instanton term in the  mass matrix (\ref{QuarkMasses}),
all the fundamental fields remain massive despite $\omega_-=0$, so the low-energy
degree of freedom are comprised of the gauge and bifundamental fields.
Or rather, these are the dominant degrees of freedom in an infinite quiver.
In a finite quiver with boundaries, the extra `quark' and `antiquark' fields at
the boundaries are also very important.


\subsection{Deconstructing the Boundaries of the $E_0$}
In the Ho\v{r}ava--Witten orbifold context, the $x^{10}$ dimension has boundaries
where the $SU(3)_1\subset E8_1$ (at the left boundary) and the
$SU(3)_2\subset E8_2$ (at the right boundary) act as flavor symmetries
of the $E_0$ theory.
Although these $SU(3)_1\times SU(3)_2$ symmetries are gauged,  their couplings
become weak (from the 4D point of view) when the volume of the $T^6/\IZ_3$
orbifold becomes large.
Since we want to focus on the single fixed point rather than on the entire orbifold,
we take the infinitely large 6--volume limit, and in that limit we may treat
the $SU(3)_1\times SU(3)_2$ flavor symmetry of the $E_0$ boundaries as global
rather than gauged.
Thus, in the finite quiver which deconstructs the $E_0$ with boundaries, each boundary
should have a global $SU(3)$ symmetry.

Moreover, according to Ganor and Sonnenschein \cite{GanorSon02}, on the
Coulomb branch of the $E_0$,
the $SU(3)\times SU(3)$ global symmetry should be spontaneously broken down
to the $SU(2)\times SU(2)\times U(1)$.
Or taking into account the $U(1)$ gauge symmetry in the bulk which is
Higgsed down by the squark VEVs at the boundaries,
\be
\rm
SU(3)_{left\, end}\times SU(3)_{right\,end}\times U(1)_{bulk}\
\to\ SU(2)_{left}\times SU(2)_{right}\times U(1)_{left+right+bulk}\,.
\label{SymBreak}
\ee
The Higgs regime of the quiver which deconstructs the $E_0$ with boundaries should
faithfully reproduce this symmetry breaking pattern.
It should also not yield any massless 4D particles besides the 5 twisted states
of the orbifold which remain massless when the fixed-point singularity is blown up.

The simplest solution to these requirements is the following quiver:
\be
\newif\iffirstnode
\newif\iflastnode
\psset{unit=1cm,linewidth=1.2pt,arrowscale=1.5}
\def\node #1(#2){%
	\rput(#2){%
		\pscircle*[linecolor=green](0,0){0.4}
		\rput(0,0){2}
		\iffirstnode
			\psline[linecolor=blue]{->}(-1.6,0)(-0.4,0)
			\uput[d](-1,0){$Q$}
		\else
			\psline[linecolor=red]{<-}(0,+0.4)(0,+1.2)
			\uput[r](0,+1.1){$q_{\scriptscriptstyle{#1}}^{}$}
		\fi
		\psline[linecolor=blue]{->}(0.4,0)(1.6,0)
		\iflastnode
			\uput[d](1,0){$\tilde Q$}
		\else
			\uput[u](1,0){$\Omega_{\scriptscriptstyle{#1}}^{}$}
			\psline[linecolor=red]{->}(0,-0.4)(0,-1.2)
			\uput[r](0,-0.9){$\tilde q_{\scriptscriptstyle{#1}}^{}$}
		\fi
		}
	}
\begin{pspicture}(-2.4,-1.5)(11.4,+1.5)
\firstnodetrue
\lastnodefalse
\psframe*[linecolor=yellow](-2.4,-0.4)(-1.6,+0.4)
\rput(-2,0){3}
\node1(0,0)
\firstnodefalse
\node2(2,0)
\node3(4,0)
\psline[linecolor=blue,linestyle=dotted](5.6,0)(6.6,0)
\node{N-1}(7,0)
\lastnodetrue
\node{N}(9,0)
\psframe*[linecolor=yellow](10.6,-0.4)(11.4,+0.4)
\rput(11,0){3}
\end{pspicture}
\label{BigQuiver}
\ee
with extra singlets $s_\ell$, $S_i$, and $\tilde S_i$ ($i=1,2,3$)
and the superpotential
\be
W\
=\,\sum_{\ell=1}^{N-1}\Bigl(
	s_\ell\,\Omega_\ell^2\,+\,\tilde q_\ell^{}\,\Omega_\ell\,q_{\ell+1}^{}\Bigr)\
+\ \epsilon^{ijk}S_iQ_jQ_k\ +\ \epsilon^{ijk}\overline S_i\tilde Q_j\tilde Q_k
\label{BigW}
\ee
Note the $SU(3)_1\times SU(3)_2$ flavor symmetry of this superpotential:
the $SU(3)_1$ symmetry acts on the $Q_i$  the $S_i$ fields at the left end of the quiver,
while the $SU(3)_2$ acts on the $\tilde Q_i$ and the $\tilde S_i$ at the right end.

The simplest way to obtain this theory is to start with
the infinite $SU(2)$ quiver (\ref{SU2quiver}) and the superpotential~(\ref{SU2quiverW}),
turn off the gauge fields of the $\ell=0$ and $\ell=N+1$ nodes, and throw
away all chiral and gauge fields which decouple from the $\ell=1,\ldots,N$ nodes.
The surviving fields left of the $\ell=1$ node comprise the $\Omega_0$  (which becomes
two $SU(2)_1$ doublets), the $\tilde q_{-1}$ (which becomes two singlets), and the singlet $s_0$.
Together with the $q_0$ doublet, they become three doublets $Q_i$ plus three singlets $S_i$
with the Yukawa couplings between them amounting to $W\supset\epsilon^{ijk}S_iQ_jQ_k$.
Likewise, the surviving fields to the right of the $\ell=N$ node comprise the $\Omega_N$,
the $q_{N+1}$ and the $s_0$, which together with the $\tilde q_N$ become three $SU(2)_N$
doublets $\tilde Q_i$ and three singlets $\tilde S_i$, with the Yukawa couplings
$W\supset \epsilon^{ijk}\overline S_i\tilde Q_j\tilde Q_k$ to each other.

Now consider the Higgs regime of the quiver (\ref{BigQuiver}).
The Yukawa couplings to the singlets restrict the scalar VEV matrices $\vev{\Omega_\ell}$,
$\vev{Q}$ and $\vev{\tilde Q}$ to $\rm rank\le1$, while the $[SU(2)]^N$ D-terms
require all these VEVs to have similar magnitudes.
Thus, modulo gauge and flavor symmetries of the theory, the only flat direction
of the scalar potential of the quiver is
\be
{\rm all}\quad\vev{Q_\ell}\,=\,\begin{pmatrix} \omega & 0 \\ 0 & 0 \\ \end{pmatrix},\qquad
\vev{Q}\,=\,\begin{pmatrix} \omega & 0 \\ 0 & 0 \\ 0 & 0 \\ \end{pmatrix},\qquad
\vev{\tilde Q}\,=\,\begin{pmatrix} \omega & 0 & 0 \\ 0 & 0 & 0 \\ \end{pmatrix},
\label{HiggsVEVs}
\ee
with the same modulus $\omega$ governing all these VEVs.
For $|\omega|>|\Lambda|$ we expect the quiver to be in the Higgs regime,
so the semiclassical analysis should adequately describe the low-energy physics.
Here are the highlights:
\begin{itemize}
\item
The entire $[SU(2)]^N$ gauge theory of the quiver is Higgsed down to nothing.%
\footnote{%
	For a long quiver, the $[U(1)]^N$ photons have a Kaluza-Klein-like
	tower of light modes with $O(1/(Na))$ masses, but there is no zero mode
	due to Higgsing by $\vev{Q}$ and $\vev{\tilde Q}$ at the ends of the quiver.
	For the charged $W^\pm$ gauge fields, all the modes have heavy $O(\omega)$ masses.
	}

\item
Most of the chiral superfields of the theory that are not eaten by the Higgs mechanism
get $O(\omega)$ masses from the the superpotential~(\ref{BigW}) or $O(\Lambda^3/\omega^2)$
masses from the one-instanton effects.

\item
The $SU(3)_1\times SU(3)_2$ global symmetry of the theory is spontaneously broken
down to $SU(2)_1\times SU(2)_2\times U(1)_{\rm combined}$, in perfect agreement with
Ganor and Sonnenschein.

\item
The only massless particles are the 9 Goldstone bosons of the global symmetry breakdown
and their superpartners, packaged into 5 chiral multiplets, namely:
\begin{itemize}
\item
The modulus $\omega$ or rather its variation $\delta\omega$, which affects
the $Q_{1,(1)}$, the $\tilde Q_1^{(1)}$, and all the $\Omega_{\ell,(1)}^{(1)}$ fields.
\item
The quarks $Q_{2,(1)}$ and $Q_{3,(1)}$.
\item
The antiquarks $\tilde Q_2^{(1)}$ and $\tilde Q_3^{(1)}$.
\end{itemize}

\item
The $x^{10}$ locations of these particles are precisely as in Ganor and Sonnenschein:
\be
\vcenter{%
	\setbox0=\hbox{{\darkgreen bulk+boundaries} $\longrightarrow$}
	\setbox2=\hbox to \wd0 {\hfil {\blue left boundary} $\longrightarrow$}
	\box0
	\vskip 0.5\baselineskip
	\box2
	\vskip 0.5\baselineskip
	}\enspace
\begin{pmatrix}%
	\darkgreen\delta\omega & \blue\overline Q_2^{(1)} & \blue\overline Q_3^{(1)}\cr
	\blue Q_2^{(1)} &  \red * & \red *\cr
	\blue Q_3^{(1)} & \red * & \red *\cr
\end{pmatrix}\enspace
\vcenter{%
	\setbox0=\hbox{$\longleftarrow$ {\blue right boundary}}
	\setbox2=\hbox{$\longleftarrow$ {\red missing}}
	\box0
	\vskip 0.5\baselineskip
	\box2
	\vskip 0.5\baselineskip
	}
\ee

\item[$\star$]
To summarize, the low-energy physics of the Higgs regime of the quiver (\ref{BigQuiver})
is in good agreement with the Coulomb branch of the $E_0$ theory with boundaries.

\end{itemize}

\subsection{The Confinement Regime}
Now consider the confinement regime of the quiver~(\ref{BigQuiver}),
which corresponds to the superconformal point of the $E_0$.
For the infinite quiver, this regime obtains for any $|\omega|<|\Lambda|$,
so for simplicity's sake, let's assume $\omega=0$, $i.\,e.$, no scalar VEVs
whatsoever.
We shall return to the effects of $\omega\neq0$ on
a finite-length quiver in the next subsection \S4.4.

Note that each $SU(2)_\ell$ factor of the quiver~(\ref{BigQuiver}) couples to 6 doublets,
so it acts as SQCD with $n_c=2$ colors and $n_f=3=n_c+1$ flavors.
The IR behavior of such theories is confinement without chiral symmetry breaking.
Instead, there are massless composite particles --- the mesons and the baryons.
However, tree-level Yukawa couplings of the quarks and antiquarks to  singlets
(or more general, to fields not charged under the $SU(2)_\ell$ in question) would render
some of the mesons and the baryons massive, and the singlets would also become massive.
We shall see momentarily that for the quiver~(\ref{BigQuiver}), all the fundamental
and the bifundamental fields become confined, while most of the composite particles
and all the singlets become massive due to Yukawa couplings.
The only particles which remain exactly massless are the 9 meson-like states
comprising a quark at one end of the quiver, an antiquark at the other end,
and all of the bifundamental fields,
\be
M_{ij}\ =\ \Lambda^{-N}\bigl(Q_i\Omega_1\Omega_2\cdots\Omega_{N-1}\tilde Q_j\bigr).
\label{Mesons}
\ee
The $SU(3)_1\times SU(3)_2$ flavor symmetry of the quiver remains unbroken
in the confinement regime,
and the mesons~(\ref{Mesons}) form the  $({\bf 3},{\bf\bar3})$ multiplet of this
symmetry.
This is in perfect agreement with the heterotic twisted states $T_{ij}$ at the un-blow-up
fixed point.
Also, we shall see that the mesons~(\ref{Mesons}) have Yukawa couplings to each other
of the form $W\supset\det(M_{ij})$, exactly as the heterotic twisted states $T_{ij}$.

To see how it works, let's start with a warm-up exercise of the quiver of length $N=1$:
A single $SU(2)$ gauge group, with 3 quarks $Q_i$, 3 antiquarks $\tilde Q_i$, 6 singlets
$S_i$ and $\tilde S_i$, and the Yukawa couplings
\be
W\ =\ \epsilon^{ijk}S_iQ_jQ_k\ +\ \epsilon^{ijk}\overline S_i\tilde Q_j\tilde Q_k\,.
\label{OneYukawa}
\ee
This theory confines without chiral symmetry breaking, and without the Yukawa couplings
it would produce 15 massless supermultiplets: 9 mesons $M_{ij}=Q_i\tilde Q_j/\Lambda$,
3 baryons $B^i=\epsilon^{ijk}Q_jQ_k/\Lambda$, and 3 antibaryons
$\tilde B^i=\epsilon^{ijk}\tilde Q_j\tilde Q_k/\Lambda$.
But the tree-level Yukawa couplings~(\ref{OneYukawa}) become mass terms
for the baryons, antibaryons, and all the singlets, so only the nine mesons $M_{ij}$
remains massless.

These mesons have non-perturbative Yukawa couplings to each other,
\be
W_{\rm NP}\ \sim\ \det(M)\ \sim\ \epsilon^{ijk}\epsilon^{lmn}M_{il}M_{jm}M_{kn}\,.
\ee
They also have Yukawa couplings to the baryons and to other massive particles,
but for the present purposes we shall focus on the couplings among the massless
particles only.

Now consider a more involved example of the two-node quiver, $N=2$:
\begin{gather}
\psset{linewidth=1.2pt,arrowscale=1.5}
\begin{pspicture}[shift=-1.4](-2.4,-1.5)(+5,+1.2)
\psframe*[linecolor=yellow](-2.4,-0.4)(-1.6,+0.4)
\rput(-2,0){3}
\psframe*[linecolor=yellow](+3.6,-0.4)(+4.4,+0.4)
\rput(+4,0){3}
\pscircle*[linecolor=green](0,0){0.4}
\rput(0,0){2}
\pscircle*[linecolor=green](2,0){0.4}
\rput(2,0){2}
\psline[linecolor=blue]{->}(-1.6,0)(-0.4,0)
\uput[d](-1,0){$Q$}
\psline[linecolor=blue]{->}(+0.4,0)(+1.6,0)
\uput[u](+1,0){$\Omega$}
\psline[linecolor=blue]{->}(+2.4,0)(+3.6,0)
\uput[d](+3,0){$\overline Q$}
\psline[linecolor=red]{->}(0,-0.4)(0,-1.2)
\uput[r](0,-0.9){$\bar q$}
\psline[linecolor=red]{->}(2,+1.2)(2,+0.4)
\uput[r](2,+0.9){$q$}
\end{pspicture}
+\ {\rm singlets}\ S_i,\overline S_i,s\\
W\ =\ \epsilon^{ijk}S_iQ_jQ_k\
+\ \epsilon^{ijk}\overline S_i\overline Q_j\overline Q_k\
+\ s\,\Omega^2\ +\ \bar q\,\Omega\, q .
\end{gather}
For the moment, let's give un-equal gauge couplings to the two $SU(2)$ gauge groups
of the quiver so that $\Lambda_1\gg\Lambda_2$.
In this case, the non-perturbative effects of the $SU(2)_1$ group are felt at higher energies,
so we may focus on the $SU(2)_1$ non-perturbative effects first, truncate the resulting
particle spectrum to the massless particles only, and only then couple them to the $SU(2)_2$.
Thus, from the $SU(2)_1$ point of view,
the $\tilde q^{(\alpha)}$ and the $\Omega^{(\alpha)}_{\,\,(\beta)}$ are 3 antiquarks doublets
while the $s$ and the $q_{(\beta)}$ are 3 singlets coupled to those antiquarks,
just like the $\tilde S_i$ couple to the $\tilde Q_i$ in the single-node example.
Thus, when the $SU(2)_1$ confines, it makes massless mesons
\be
P_{i,(\beta)}\ =\ {1\over\Lambda_1}\,Q_i\Omega_{(\beta)}\quad{\rm and}\quad
R_i\ =\ {1\over\Lambda_1}\,Q_i\tilde q
\ee
but the baryons $Q_{[i}Q_{j]}$, the antibaryons $\Omega^2$ and $\tilde q\Omega_{(\beta)}$
and the singlets $S_i$, $s$, and $q^{(\beta)}$ become massive.

Now from the $SU(2)_2$  point of view,
the $P_{i,(\beta)}$ mesons are 3 doublets, so they act as quarks,
while the $R_i$ mesons act as 3 singlets.
Combining these fields with the antiquarks $\tilde Q_i^\beta$ and singlets $\tilde S_i$,
we end up with the SQCD with 2 colors, 3 flavors, 6 singlets, and the Yukawa couplings
\be
W\ =\ \epsilon^{ijk}R_iP_jP_k\ +\ \epsilon^{ijk}\overline S_i\tilde Q_j\tilde Q_k\,,
\label{TwoYukawa}
\ee
where the first term in the non-perturbative 3-meson coupling of the first $SU(2)$
while the second term is tree-level.
Altogether, we get a theory exactly similar to the single-node example, so it behaves
in the same way: confines the quarks and the antiquarks
without breaking the $SU(3)\times SU(3)$ chiral symmetry,
makes massless mesons
\be
M_{ij}\ =\ {1\over\Lambda_2}\,P_i\tilde Q_j\
=\ {1\over\Lambda_1\Lambda_2}\,Q_i\Omega\tilde Q_j\,,
\label{DoubleMesons}
\ee
while the baryons, the antibaryons, and the singlets $R_i$ and $\tilde S_i$ become massive.

Now suppose $\Lambda_2\gg\Lambda_1$ instead of the other way around.
In this case, the $SU(2)_2$ confines first, makes massless mesons which from the
$SU(2)_1$ point of view look like 3 antiquarks plus 3 singlets, and then the confinement
in the $SU(2)_1$ makes massless meson-like particles exactly as in eq.~(\ref{DoubleMesons}).
So, the quivers with $\Lambda_1\gg\Lambda_2$ and with $\Lambda_2\gg\Lambda_1$
have exactly the same massless composite particles, made from exactly the same
quark, bifundamental, and antiquark fields, and with the same Yukawa couplings
$W\supset\det(M_{ij})$ to each other, while every other particle in the theory
--- elementary or composite --- becomes massive.

Based on this complementarity, we believe that two-node quivers with all
$\Lambda_1/\Lambda_2$ ratios have the same spectrum of massless particles, namely
the 3 quark-bifundamental-antiquark composites~(\ref{DoubleMesons}).
In particular, such mesons should be the only massless particles for
the quiver with $\Lambda_1=\Lambda_2$.

Generalizing the above analysis to quivers~(\ref{BigQuiver})
with any numbers of nodes is completely straightforward.
For simplicity, we shall proceed by dealing with one $SU(2)_\ell$ factor at a times
as would be appropriate for $\Lambda_1\gg\Lambda_2\gg\cdots\gg\Lambda_N$,
but the massless spectrum obtaining at the end of the process should be valid
for all ratios of confinement scales, and in particularly for the
$\Lambda_1=\Lambda_2=\cdots=\Lambda_N$.
Thus we start with the confining $SU(2)_1$ which has 3 quarks, three other doublets acting
as antiquarks, and 6 singlets (or fields without $SU(2)_1$ charges), exactly as in
the two-node example, so the composite massless particles are the $P_{i,(\beta)}$
(which act as 3 quarks of the $SU(2)_2$) and the singlets $R_i$,
while the baryons, the antibaryons, and the elementary $S_i$, $\sigma$,
and $q_{2,(\beta)}$ fields become massive.
Consequently, we end up with the quiver
\be
\newif\iffirstnode
\newif\iflastnode
\psset{unit=1cm,linewidth=1.2pt,arrowscale=1.5}
\def\node #1(#2){%
	\rput(#2){%
		\pscircle*[linecolor=green](0,0){0.4}
		\rput(0,0){2}
		\iffirstnode
			\psline[linecolor=blue]{->}(-1.6,0)(-0.4,0)
			\uput[d](-1,0){$P$}
		\else
			\psline[linecolor=red]{<-}(0,+0.4)(0,+1.2)
			\uput[r](0,+1.1){$q_{\scriptscriptstyle{#1}}^{}$}
		\fi
		\psline[linecolor=blue]{->}(0.4,0)(1.6,0)
		\iflastnode
			\uput[d](1,0){$\tilde Q$}
		\else
			\uput[u](1,0){$\Omega_{\scriptscriptstyle{#1}}^{}$}
			\psline[linecolor=red]{->}(0,-0.4)(0,-1.2)
			\uput[r](0,-0.9){$\tilde q_{\scriptscriptstyle{#1}}^{}$}
		\fi
		}
	}
\begin{pspicture}(-0.4,-1.5)(11.4,+1.5)
\firstnodetrue
\lastnodefalse
\psframe*[linecolor=yellow](-0.4,-0.4)(+0.4,+0.4)
\rput(0,0){3}
\node2(2,0)
\firstnodefalse
\node3(4,0)
\psline[linecolor=blue,linestyle=dotted](5.6,0)(6.6,0)
\node{N-1}(7,0)
\lastnodetrue
\node{N}(9,0)
\psframe*[linecolor=yellow](10.6,-0.4)(11.4,+0.4)
\rput(11,0){3}
\end{pspicture}
\ee
which looks exactly like the original quiver~(\ref{BigQuiver}) except that
it is shorter by one node.

At this point we repeat the procedure focusing on the confining $SU(2)_2$,
and in the same manner end up with a quiver of length $N-2$, {\it etc., etc.}
Eventually, we arrive at a single-node quiver, and after dealing with the
confining $SU(2)_N$, we end up with nine massless meson-like states
\be
M_{ij}\ =\ {1\over\Lambda_1\cdots\Lambda_N}\,
\bigl(Q_i\Omega_1\Omega_2\cdots\Omega_{N-1}\tilde Q_j\bigr)
\label{BigMesons}
\ee
while every particle --- elementary or composite --- is massive.

Note that while it is much simpler to deal with one confining $SU(2)$ factor
at a time, we can handle them in any order we like.
In particular, we may start in the middle of the quiver and work our way outwards,
or even jump around the quiver to non-adjacent nodes in a random fashion.
While the technical details of such random-order procedure are too boring
to be presented here, let us simply state the bottom line:
regardless of the order in which we handle the $N$ $SU(2)_\ell$ factors,
we always end up wit the same 9 massless particles (\ref{BigMesons}).
Consequently, we belie that the same 9 massless particles emerge for any
ratios of the confinements scales $\Lambda_\ell$, including the equal-scales
case of $\Lambda_1=\Lambda_2=\cdots=\Lambda_N$.

So let us re-iterate the bottom line of this subsection:
The massless twisted states $T_{ij}$ at an un-blown fixed point of the $T^6/\IZ_3$
orbifold --- or rather of the Ho\v{r}ava--Witten dual of the orbifold ---
deconstruct to the massless meson-like states~(\ref{BigMesons}) of the quiver
(\ref{BigQuiver}) in its confinement regime.

\subsection{Deconstructing the Blow Up}
In the heterotic string theory, blowing up a fixed point is parametrized
by the VEVs $\vev{T_{ij}}$ of  twisted-sector scalars.
Up to an $SU(3)\times SU(3)$ symmetry, the VEV matrix looks like
\be
\vev{T_{ij}}\ =\,\begin{pmatrix} t & 0 & 0 \\ 0 & 0 & 0\\ 0 & 0 & 0 \\ \end{pmatrix}
\label{BlowUp}
\ee
and the $W\supset\det(T)$ Yukawa couplings give ${\rm mass}=t$
to the 4 of the $T_{ij}$ states,
while the remaining 5 states remain massless.
For the Ho\v{r}ava--Witten dual of the blown-up orbifold, Ganor and Sonnenschein
have identified the 5 massless  states $T_{11},T_{12},T_{13},T_{21},T_{31}$
of the Coulomb-branch $E_0$ theory with boundaries,
but they saw no sign of the 4 massive states $T_{22},T_{23},T_{32},T_{33}$.
However, for small $t$ --- $i.\,e.$, when the fixed point is just a little bit
blown up --- the $T_{22},T_{23},T_{32},T_{33}$ states become very light
so they should be easy to identify.
In this subsection, we resolve this paradox for the deconstructed $E_0$.

Consider the Coulomb-branch moduli of the quiver~(\ref{BigQuiver}).
In terms of independent holomorphic gauge-invariant combinations of the chiral
superfields, there are 9 such moduli, namely the
\be
{\cal M}_{ij}\ =\ Q_i\Omega_1\Omega_2\cdots\Omega_{N-1}\tilde Q_j\,,\quad
i,j=1,2,3,
\label{Qmoduli}
\ee
subject to the $\mathop{\rm rank}({\cal M})=1$ constraint due to superpotential
\be
W({\cal M})\ \sim\ {1\over\Lambda^{3N}}\,\det({\cal M}).
\label{WM}
\ee
At the origin ${\cal M}_{ij}=0$ of the moduli space, all 9 moduli (\ref{Qmoduli})
give rise to massless particles $T_{ij}$, so the moduli space metric $g_{\cal\bar M M}$
should be non-singular at the origin.
In terms of the K\"ahler function of the moduli space, this means
\be
K\bigl(\overline{\cal M},{\cal M}\bigr)\
=\ {1\over|\Lambda|^{2N}}\,\tr\bigl(\overline{\cal M}{\cal M}\bigr)\
+\ O\left({\overline{\cal M}{\cal M}\overline{\cal M}{\cal M}\over|\Lambda|^{4N+2}}\right)
\label{KMC}
\ee
A {\it small} blow-up (\ref{BlowUp}) of the fixed point corresponds to
\be
\vev{{\cal M}_{11}}\ =\ \Lambda^N\times t,\quad t\ll\Lambda,
\ee
which gives the twisted-sector particles $T_{22},T_{23},T_{32},T_{33}$ {\it physical} masses
\be
m\ =\ |\vev{{\cal M}_{11}}|\times
{\rm Yukawa\,coupling\,from\,(\ref{WM})\over {\mit g}_{\cal\bar M M}\,from\,(\ref{KMC})}\
=\ |t|.
\ee
As long as $t\ll\Lambda$, this mass remains much lighter than the $O(\Lambda)$
masses of all the other massive particles of the quiver theory.

In terms of the $\omega$ parameter of the scalar VEVs (\ref{HiggsVEVs}),
\be
\vev{{\cal M}_{11}}\ =\ \omega^{N+1}
\quad\Longrightarrow\quad
t\ =\ {\omega^{N+1}\over\Lambda^N}\,.
\ee
Hence, in the long quiver limit $N\to\infty$, any $\omega<\Lambda$ corresponds
to a very small $t\ll\Lambda$.
Physically, this means negligible blow-up and hence negligible masses of the
$T_{22},T_{23},T_{32},T_{33}$ twisted particles.
{\it This is why for a long quiver, any scalar VEV $\omega<\Lambda$ is as good as
$\omega=0$:\/} the confinement regime is un-affected and the fixed point
of the deconstructed orbifold remains un-blown.

Now consider the Higgs regime of the quiver with $\omega>\Lambda$,
which corresponds to very large $\vev{{\cal M}_{11}}$ and $t\gg\Lambda$.
In this regime, the K\"ahler function of the the quiver's moduli space is quite
different from  eq.~(\ref{KMC}).
From the semiclassical Higgs fields~(\ref{HiggsVEVs}), we expect
\be
K\ \approx\ (N+1)\overline\omega\omega\,,
\ee
so re-expressing $K$ in terms of ${\cal M}_{ij}$ and $\overline{\cal M}_{ij}$
and requiring the $SU(3)\times SU(3)$ symmetry gives us
\be
K\ \approx\ (N+1)\root N+1\of{\tr\bigl(\overline{\cal M}{\cal M}\bigr)}.
\label{KMH}
\ee
Consequently, the K\"ahler metric for the fields
${\cal M}_{22},{\cal M}_{23},{\cal M}_{32},{\cal M}_{33}$ is
\be
g_{\overline{\cal M}{\cal M}}\ =\ {1\over|\omega|^{2N}}\,,
\ee
hence the physical mass of the $T_{22},T_{23},T_{32},T_{33}$ twisted particles is
\be
m\ =\ \left|{\omega^{3N+1}\over\Lambda^{3N}}\right|\ \gg\ \Lambda\,.
\ee
This extremely large mass explains why we do not see these particles
in the Ganor--Sonnenschein construction of the blown-up fixed point
of the Ho\v{r}ava--Witten orbifold.

We hope the above arguments explain the abrupt transition at $\omega=\Lambda$
between the confinement and the Higgs regimes in the $N\to\infty$ limit of
the quiver~(\ref{BigQuiver}).
In $E_0$ terms, this is the transition between the superconformal theory
on the un-blown fixed plane and the Coulomb-branch $U(1)_{\rm 5D}$ theory
on the blown-up plane.
For the quivers of large but finite length $N$, the transition takes a finite
but narrow range of scalar VEVs $\omega$,
\be
1\ -\ O\left({1\over N}\right)\ <\ {\omega\over\Lambda}\ <\ 1\ +\ O\left({1\over N}\right).
\ee
Thus, for the Ho\v{r}ava--Witten theory with a finite length of $x^{10}$
--- which corresponds to a large but finite heterotic string coupling, ---
we expect a  sharp but continuous  transition between the Calabi--Yau regime
of a blown-up fixed point and some non-geometric regime
hiding behind the un-blown fixed point.

The similar sharp crossover is well known for many 5D theories (including the $E_0$)
compactified on a large circle.
For the 5D theories compactified on the interval with boundaries, the general
behavior should be similar, but the details need to be worked out.
We hope the present paper sheds some light on the $E_0$ theory on the interval.

\section{Work in Progress and Open Questions}
\renewcommand{\labelitemi}{$\star$}
\renewcommand{\labelitemii}{$\bullet$}
Through most of this paper we have focused on deconstructing
the twisted states of a particular $T^6/\IZ_3$ orbifold.
The obvious next step is to apply the same method to other heterotic
orbifolds: Deconstruct the 5D SCFT at each fixed plane, work out the
boundary `quarks' and incorporate them into the quiver, and to see if the
confining regime of the quiver indeed produces massless meson-like states
with quantum numbers matching the twisted states of the heterotic string.
This work is in progress: Thus far, we have worked out a few $T^6/\IZ_4$
and $T^6/\IZ_6$ models \cite{Jacob16}; the boundary `quarks' and hence the
quiver's ends in these models are more complicated than in the $T^6/\IZ_3$
model, but the non-local massless mesons of the quivers do match the massless
twisted states of the corresponding orbifold.
We hope to work out a few more models to see how the quiver's ending depend
on a particular model before we present all the deconstructed orbifolds in
a separate paper.

But besides the technical issues of the quiver boundaries, the very fact that in
the Ho\v{r}ava--Witten theory the twisted states become non-local `mesons'
spanning the entire length of the $x^{11}$ dimensions raises a several deep
questions:
\begin{itemize}
\item
First of all, what is the physical meaning of the gluon string connecting
the quark at one end of the $x^{10}$ to the antiquark at the other end?
In the deconstructed theory, this string is the product of all the bifundamental
scalar fields of the quiver, but what does it become in the continuum limit?
A flux tube?  A Wilson line? Something else?

\item
Second, what is the M-theory origin of this gluon string?
It does {\it not} look like an M2 or M5 brane wrapped around some cycle
of the $\IC^3/\IZ_3$ singularity, so what else can it be?

\item
Regardless of the gluon string's origin, it is tensionless:
It's the only way to keep the twisted states including this string massless
in the long $x^{10}$ regime dual to the strong heterotic coupling.
So what happens when the tensionless string becomes long?
Does it run straight from one end of the $x^{10}$ to the other end,
or can it wiggle in the $x^{1,2,3}$ directions of the ordinary 3D space?
Can the quark at one end of the string move away
(in the $x^{1,2,3}$ directions) from the antiquark at the other end?
If yes, does it mean that the twisted particles become `fat' rather than
nearly-pointlike in the strong heterotic coupling regime?

\begin{itemize}
\item
Tentatively, the answers to these questions depend on the higher-derivative
terms in the world-sheet Lagrangian for the  gluon string.
If all such terms vanish with the tension, then the string can wiggle as much as it wants,
the quarks and the antiquark can separate in 3D space,
and the twisted-sector particles have effective size comparable to the length
of the $x^{11}$ (since this is the only scale of the locally-conformal 5D theory
on the fixed plane).
But if the higher-derivative terms do not vanish, then they stiffen the string
and might force it to run in a straight line in the $x^{10}$ direction, which in turn
would keep the quark and the antiquark from separating in $x^{1,2,3}$ directions.

Alas, without knowing the nature of the string we cannot say if it has
any higher-derivative terms or not.

\item
Without delving into the nature of the gluon string, can we probe
for the quark-antiquark separation through some gedankenexperiment?
What would be a good probe of such separation?
The form-factor in some scattering process involving
both the $SU(3)_1$ and the $SU(2)_2$ charges of the twisted states?
Something else?

\end{itemize}
\end{itemize}

\acknowledgments
The research on which this article is based was partially supported by the
US National Science Foundation under grant 1417366 (both authors) and by the
US--Israel Bi-National Science Foundation (VK).


\end{document}

\end{document}